\documentclass[preprint2,tighten]{aastex6}

\newcommand{\dustings}{DUSTiNGS}
\newcommand{\spitzer}{{\it Spitzer}}
\newcommand{\hst}{{\it HST}}

\usepackage{enumitem}

\usepackage{amsmath}

\usepackage{subfigure}
\usepackage{graphicx}

\usepackage{ascii}
\usepackage[T1]{fontenc}

\begin{document}

\title{An Infrared Census of DUST in Nearby Galaxies with Spitzer (DUSTiNGS). IV. Discovery of High-Redshift AGB Analogs}

\author{M.~L.\ Boyer\altaffilmark{1}$^{\star}$}
\and
\author{K.~B.~W.\ McQuinn\altaffilmark{2}}
\and
\author{M.~A.~T.\ Groenewegen\altaffilmark{3}}
\and
\author{A.~A.\ Zijlstra\altaffilmark{4,5}}
\and
\author{P.~A.\ Whitelock\altaffilmark{6,7}}
\and
\author{J.~Th.\ van~Loon\altaffilmark{8}}
\and
\author{G.\ Sonneborn\altaffilmark{9}}
\and
\author{G.~C.\ Sloan\altaffilmark{1,10}}
\and
\author{E.~D.\ Skillman\altaffilmark{11}}
\and
\author{M.\ Meixner\altaffilmark{1}}
\and
\author{I.\ McDonald\altaffilmark{4}}
\and
\author{O.\ Jones\altaffilmark{1}}
\and
\author{A.\ Javadi\altaffilmark{12}}
\and
\author{R.~D.\ Gehrz\altaffilmark{11}}
\and
\author{N.\ Britavskiy\altaffilmark{13}}
\and
\author{A.~Z.\ Bonanos\altaffilmark{14}}

  \altaffiltext{1}{STScI, 3700 San Martin Drive, Baltimore, MD 21218 USA}
  \altaffiltext{2}{University of Texas at Austin, McDonald Observatory, 2515 Speedway, Stop C1402, Austin, Texas 78712 USA}
  \altaffiltext{3}{Royal Observatory of Belgium, Ringlaan 3, B-1180 Brussels, Belgium}
  \altaffiltext{4}{Jodrell Bank Centre for Astrophysics, Alan Turing Building, University of Manchester, M13 9PL, UK}
  \altaffiltext{5}{Department of Physics \& Laboratory for Space Research, University of Hong Kong, Pokfulam Rd, Hong Kong}
  \altaffiltext{6}{Astronomy Department, University of Cape Town, 7701 Rondebosch, South Africa}
  \altaffiltext{7}{South African Astronomical Observatory (SAAO), P.O. Box 9, 7935 Observatory, South Africa}
  \altaffiltext{8}{Lennard-Jones Laboratories, Keele University, ST5 5BG, UK}
  \altaffiltext{9}{Observational
    Cosmology Lab, Code 665, NASA Goddard Space Flight Center,
    Greenbelt, MD 20771 USA}
  \altaffiltext{10}{Department of Physics and Astronomy, University
of North Carolina, Chapel Hill, NC 27599-3255, USA}
  \altaffiltext{11}{Minnesota Institute for Astrophysics, School of Physics and Astronomy, 116 Church Street SE, University of
    Minnesota, Minneapolis, MN 55455 USA}
  \altaffiltext{12}{School of Astronomy, Institute for Research in Fundamental Sciences (IPM), PO Box 19395-5531, Tehran, Iran}
  \altaffiltext{13}{Instituto de Astrof\'{i}sica de Canarias, E-38205 La Laguna 38200, Tenerife, Spain}
  \altaffiltext{14}{IAASARS, National Observatory of Athens, GR-15236 Penteli, Greece}

\email{$\star$ Email: mboyer@stsci.edu}

\begin{abstract}
The survey for DUST in Nearby Galaxies with {\it Spitzer} (\dustings)
identified several candidate Asymptotic Giant Branch (AGB) stars in
nearby dwarf galaxies and showed that dust can form even in very
metal-poor systems (Z $\sim$$0.008\,Z_\odot$).  Here, we present
a follow-up survey with WFC3/IR on the {\it Hubble Space Telescope}
(HST), using filters that are capable of distinguishing carbon-rich
(C-type) stars from oxygen-rich (M-type) stars: F127M, F139M, and
F153M. We include six star-forming DUSTiNGS galaxies (NGC\,147,
IC\,10, Pegasus\,dIrr, Sextans\,B, Sextans\,A, and Sag\,DIG), all more
metal-poor than the Magellanic Clouds and spanning 1~dex in
metallicity.  We double the number of dusty AGB stars known in these
galaxies and find that most are carbon rich. We also find 26 dusty
M-type stars, mostly in IC\,10. Given the large dust excess and tight
spatial distribution of these M-type stars, they are most likely on
the upper end of the AGB mass range (stars undergoing Hot Bottom
Burning). Theoretical models do not predict significant dust
production in metal-poor M-type stars, but we see evidence for dust
excess around M-type stars even in the most metal-poor galaxies in our
sample ($12+\log({\rm O/H}) = 7.26-7.50$). The low metallicities and
inferred high stellar masses (up to $\sim$10~$M_\odot$) suggest that
AGB stars can produce dust very early in the evolution of galaxies
($\sim$30~Myr after they form), and may contribute significantly to
the dust reservoirs seen in high-redshift galaxies.
\end{abstract}


\section{INTRODUCTION}
\label{sec:intro}

The shape of the initial mass function establishes low- and
intermediate-mass stars ($\sim$0.8--10~$M_\odot$) as a dominant
contributor to the stellar populations of galaxies. These stars pass
through the thermally-pulsing Asymptotic Giant Branch
(TP-AGB)\footnote{Unless otherwise noted, we use the term AGB to
  denote TP-AGB stars throughout this paper.} phase at the end of
their evolution, during which time the products of nucleosynthesis are
returned to the interstellar medium (ISM) via a strong stellar wind
that ultimately ends the evolution of the star. The rich diversity of
elements and the dust produced during this phase make AGB stars a key
driver of galactic evolution.

Dust in high-redshift galaxies is often attributed to AGB stars and
supernovae
\citep[e.g.,][]{Valiante+2009,Dwek+2011,Rowlands+2014,Michalowski2015},
though the relative contributions from each is not well
understood. For supernovae, the uncertainty lies in the balance
between dust creation and destruction. For AGB stars, the chief
uncertainty lies in the metallicity dependence of dust production.
Some studies show a clear metallicity effect in carbon-rich AGB stars
\citep{vanLoon+2008b}, while others suggest that C star dust
production has weak-to-no metallicity dependence
\citep{Sloan+2012,Sloan+2016,Boyer+2015b}. It is, however, the dusty
      {\it M-type} AGB stars that may be more important at early times
      because the most massive AGB stars (and thus those that can
      produce dust on the shortest timescales) are O-rich. While
      dredge-up of newly-formed carbon converts AGB stars into C
      stars, a process called hot bottom burning (HBB) occurs when the
      base of the convective envelope dips into the H-shell, resulting
      in nuclear burning of $^{12}$C into $^{14}$N and thus limiting
      the formation of C stars \citep{Boothroyd+1993}. This occurs
      only in AGB stars more massive than $\sim$3 or 4~$M_\odot$,
      depending on the metallicity
      \citep{Ventura+2012,KarakasLugaro2016,Marigo+2017}. As a result,
      M-type AGB stars could inject dust into the ISM as early as
      30~Myr after forming (for a 10~$M_\odot$ star). Dusty C stars,
      on the other hand, contribute much later ($t_{\rm lifetime}
      \approx 0.3$--$3.6$~Gyr).

Unlike C stars, M-type stars do not produce their own condensable
material (typically silicon, iron, magnesium, and oxygen), so the
efficiency of dust production is expected to strongly decrease with
metallicity. This expected metallicity dependence has, however, been
difficult to quantify observationally both due to the comparative
rarity of dust-producing M-type stars and to the limited range of
metallicities reachable with IR observatories. AGB dust production in
the Magellanic Clouds ($Z/Z_\odot \simeq 0.2$ and $0.5$) has been
extensively studied \citep[e.g.,][]{Trams+1999,
  vanLoon+1998,vanLoon+2006b,vanLoon+2008b, Groenewegen+2007,
  Groenewegen+2009, Riebel+2012, Srinivasan+2016, Goldman+2017}, but
there are few examples of dust-producing AGB stars at lower
metallicities.

A handful of dusty stars have been confirmed in dwarf spheroidal
(dSph) galaxies with metallicities as low as ${\rm [Fe/H]} \sim -1$.
\citep{Sloan+2009, Sloan+2012, Lagadec+2007, Matsuura+2007,Whitelock+2009,Menzies+2010,Menzies+2011,McDonald+2014} and in globular clusters with ${\rm [Fe/H]} > -1.6$
\citep{Boyer+2008,Boyer+2009a,McDonald+2009,McDonald+2011b}. All of
these examples are C stars or low-mass M-type stars. The globular
cluster stars (low-mass M-type; $\approx$0.8--1.5~$M_\odot$) do appear
to produce dust despite their low metallicities. However, these may
not be true analogs of more massive O-rich AGB dust producers at early
epochs because (1) low-mass stars generally produce only modest
amounts of dust, (2) most are observed in globular clusters where
pollution from earlier populations is a wide-spread phenomenon
\citep{Gratton+2004,Gratton+2012,Prantzos+2007}, and (3) the mixing and
nucleosynthesis processes that occur in low-mass and high-mass O-rich
AGB stars are fundamentally different \citep{KarakasLattanzio2014}.

To search for examples of high-mass AGB dust production, environments
with large stellar populations and recent star formation must be
studied. Several dwarf galaxies in and around the Local Group are
suitable, but their large distances make IR observations difficult. A
few studies
\citep{Jackson+2007a,Jackson+2007b,Boyer+2009a,Davidge+2014,Jones+2014},
have statistically inferred the presence of dust-producing AGB stars
in several star-forming dwarf galaxies, but were unable to identify
individual dusty stars due both to confusion with unresolved
background galaxies and to substantial circumstellar extinction at
optical wavelengths. At these distances, additional information is
necessary to confidently identify individual dusty AGB
stars. \citet{McQuinn+2007} and \citet{Javadi+2013} exploited stellar
variability to identify dust-producing AGB stars in M33 and the survey
of DUST in Nearby Galaxies with
\spitzer\ \citep[\dustings;][Paper~I]{Boyer+2015a} used a similar
strategy to identify dusty AGB variables in very metal-poor
galaxies. \dustings\ observed 50 nearby galaxies and identified 526
dusty AGB candidates by their IR excesses and brightness changes
between 2 epochs, with particular sensitivity to stars with
300--600~day periods. Assuming all candidates are indeed producing
dust, \citet{Boyer+2015b} (Paper~II) found that AGB dust forms even at
metallicities as low as 0.006\,$Z_\odot$. The lack of a correlation
between dust production and metallicity in the \dustings\ galaxies
suggests that AGB stars can be a dominant source of dust even in
metal-deficient early galaxies. However, Paper~II was unable to confirm the AGB
nature of these sources or identify the AGB spectral type (and hence
whether the dust comprises silicate or carbon grains).

In this paper, we present near-IR {\it Hubble Space Telescope} (\hst)
observations of six star-forming \dustings\ galaxies. Combined with
mid-IR \spitzer\ data, the \hst\ data reveal whether the atmospheric
chemistry of the dust-producing stars is carbon- or oxygen- rich. In
addition to the confirmation of 120 dusty carbon stars, we confirm 26
dust-producing M-type stars in this sample, showing that {\it massive}
AGB stars can contribute dust at extremely low metallicity. In
Section~\ref{sec:data}, we describe the survey and stellar
classifications. In Section~\ref{sec:dust}, we discuss the properties
of the dustiest stars.

\begin{figure}
  \includegraphics[width=\columnwidth]{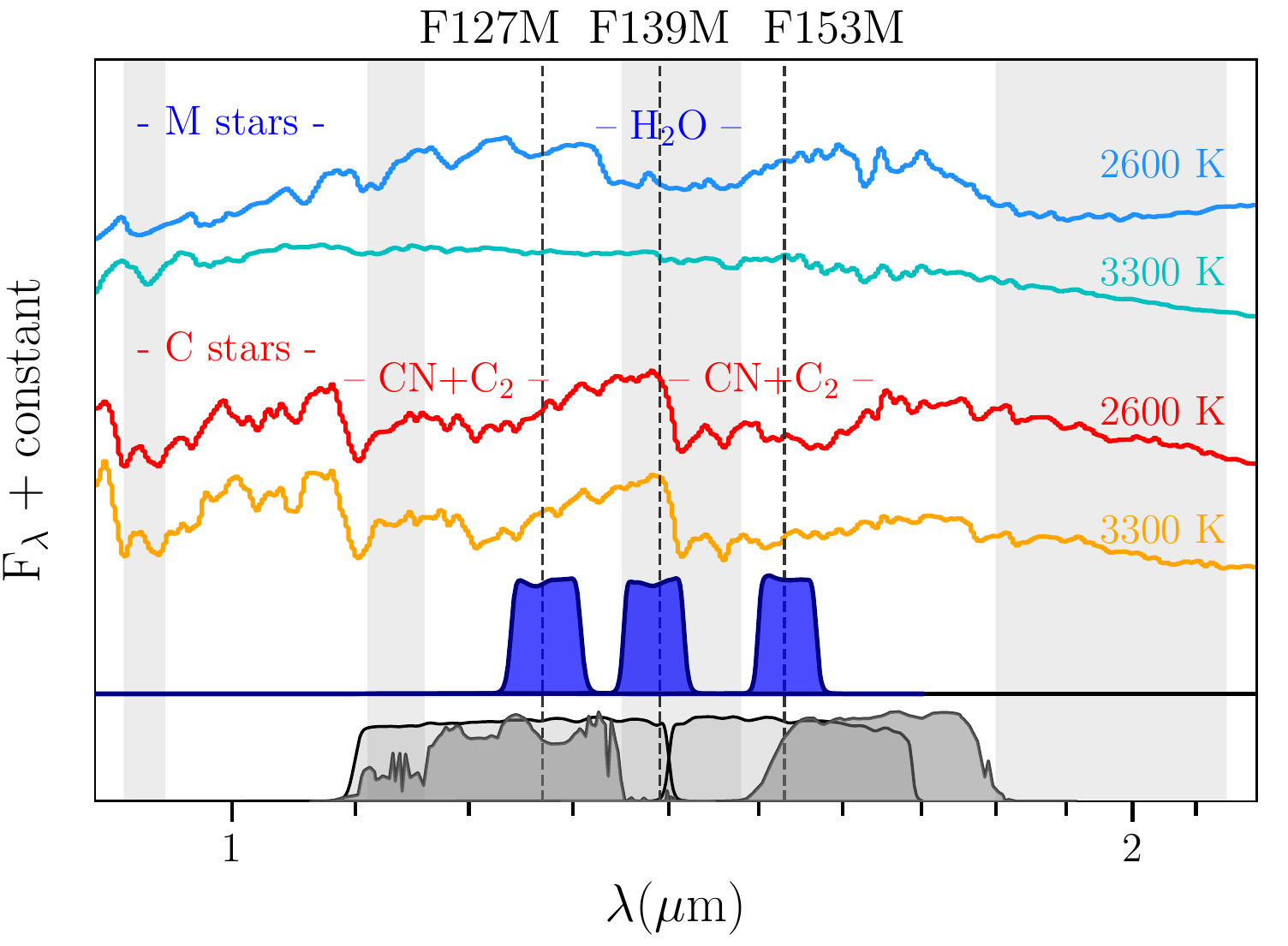}
  \caption{The WFC3/IR medium-band filters (blue) used here to
    distinguish C-type (red) and M-type (blue) stars by sampling the
    water feature in M-type stars and the CN$+$C$_2$ feature in C-type
    stars near 1.4~\micron. The model spectra are from
    \citet{Aringer+2009,Aringer+2016}. In the lower panel, we
      show the 2MASS $J$ and $H$ filters (dark gray) and WFC3/IR F125W
      and F160W filters (light gray) for reference
      (transmissions not to scale).\label{fig:filt}}
\end{figure}

\section{Data \& Analysis}
\label{sec:data}

\subsection{Identifying AGB Spectral Types with HST}

Photometric surveys typically separate C- and M-type stars using
broad-band near-IR or narrow-band optical filters. The broad-band
near-IR filters ($JHK$) are influenced by VO, TiO, and H$_2$O
molecular features in M-type stars and CN and C$_2$ in C stars, while
the narrow-band filters target TiO and CN molecular features at
$\lambda<7000$~\AA. The optical surveys are severely photon-limited
and fail to detect the stars with even moderate circumstellar dust
extinction. The near-IR $JHK$ surveys capture more of the dusty stars,
but classification is imprecise
\citep{Boyer+2013,Boyer+2015c,Ruffle+2015,Jones+2017} and the dustiest
stars, which can be faint even in the near-IR, generally remain
undetected because of source confusion and insufficient sensitivity
from the ground, especially in the $K$-band. These impediments are
overcome here with IR channel of \hst's Wide-Field Camera 3
\citep[WFC3;][]{Kimble+2008}, which has ample sensitivity and angular
resolution in the near-IR for detecting AGB stars out to the edge of
the Local Group.

Most \hst\ surveys use the wide WFC3 filters \citep[especially F110W
  and F160W;][]{Dalcanton+2012a,Dalcanton+2012b, Sabbi+2013} to
maximize the imaging depth, but these filters are too wide to be
influenced by molecular absorption features in AGB star spectra,
resulting in significant color overlap between these spectral
types. \citet{Boyer+2013} demonstrated the use of HST WFC3/IR
medium-band filters to successfully separate C- and M-type stars in a
field in the inner disk of M31. These filters fall within the H$_2$O
and CN$+$C$_2$ features at 1.2--1.5~\micron\ (filters: F127M, F139M, and
F153M, Fig.~\ref{fig:filt}), thus firmly dividing the
two spectral types with minimal cross-contamination. Furthermore, even
the dustiest stars remain separated by spectral type in the WFC3/IR
colors, unlike in $J-K$ colors which overlap significantly once dust
is included.

\begin{deluxetable*}{rrcccccc}
  \tablewidth{0pc}
  \tabletypesize{\normalsize}
  \tablecolumns{8}
  \tablecaption{Target Information\label{tab:targets}}
  \tablehead{
    \colhead{Galaxy}&
    \colhead{Alt. Name}& 
    \colhead{Morph. Type}&
    \colhead{$d$}&
    \colhead{[Fe/H]}&
    \colhead{$12+\log({\rm O/H})$}&
    \colhead{$M_{\rm V}$}&
    \colhead{$A_{\rm V}$}\\
    \colhead{}&
    \colhead{}&
    \colhead{(Mpc)}&
    \colhead{}&
    \colhead{}&
    \colhead{(mag)}&
    \colhead{(mag)}}
  \startdata
  NGC 147   & DDO 3   & dE/dSph & 0.76 & $-1.11$ & \nodata & $-14.6\pm0.1$ & 0.47 \\
  IC 10     & UGC 192 & dIrr    & 0.77 & $-1.28$ & $8.19\pm0.15$ & $-15.0\pm0.2$ & 2.33 \\
  Pegasus   & DDO 216 & dTrans/dIrr  & 0.98 & $-1.4\pm0.20$ & $7.93\pm0.13$ & $-12.2\pm0.2$ & 0.19 \\
  Sextans B & DDO 70  & dIrr    & 1.43 & $-1.6\pm0.10$ & $7.53\pm0.05$ & $-14.5\pm0.2$ & 0.09 \\
  Sextans A & DDO 75  & dIrr    & 1.46 & $-1.85$ & $7.54\pm0.06$ & $-14.3\pm0.1$ & 0.12 \\
  Sag DIG   & Sgr dIG & dIrr    & 1.09 & $-2.1\pm0.20$ & $7.26-7.50$ & $-11.5\pm0.3$ & 0.34\enddata

  \tablecomments{\ Distances are derived from F814W ($I$-band) TRGB
    estimates from \citet{McQuinn+2017}. $A_{\rm V}$ is from
    \citet{Weisz+2014}. $M_{\rm V}$ is from \citet{McConnachie2012}
    and references therein. The values of ${\rm [Fe/H]}$ adopted here
    are derived from RGB colors from \citet{Nowotny+2003},
    \citet{Bellazzini+2014}, \citet{TikhonovGalazutdinova2009},
    \citet{McConnachie+2005}, \citet{Momany+2002}, and
    \citet{Sakai+1996}. Stellar metallicities derived from
    spectroscopy of RGB stars in NGC\,147 agree with our adopted value
    \citep{Geha+2010}. However, \citet{Kirby+2017} measured ${\rm
      [Fe/H]} = -1.88^{+0.13}_{-0.09}$ from RGB star spectroscopy in
    Sag\,DIG. ISM gas-phase oxygen abundances ($12+\log({\rm O/H})$)
    are from from \citet{Mateo+1998}, \citet{HLee+2006}, and
    \citet{Saviane+2002}. }

\end{deluxetable*}

\begin{figure*}
\vbox{
\hbox{
\includegraphics[width=0.32\textwidth]{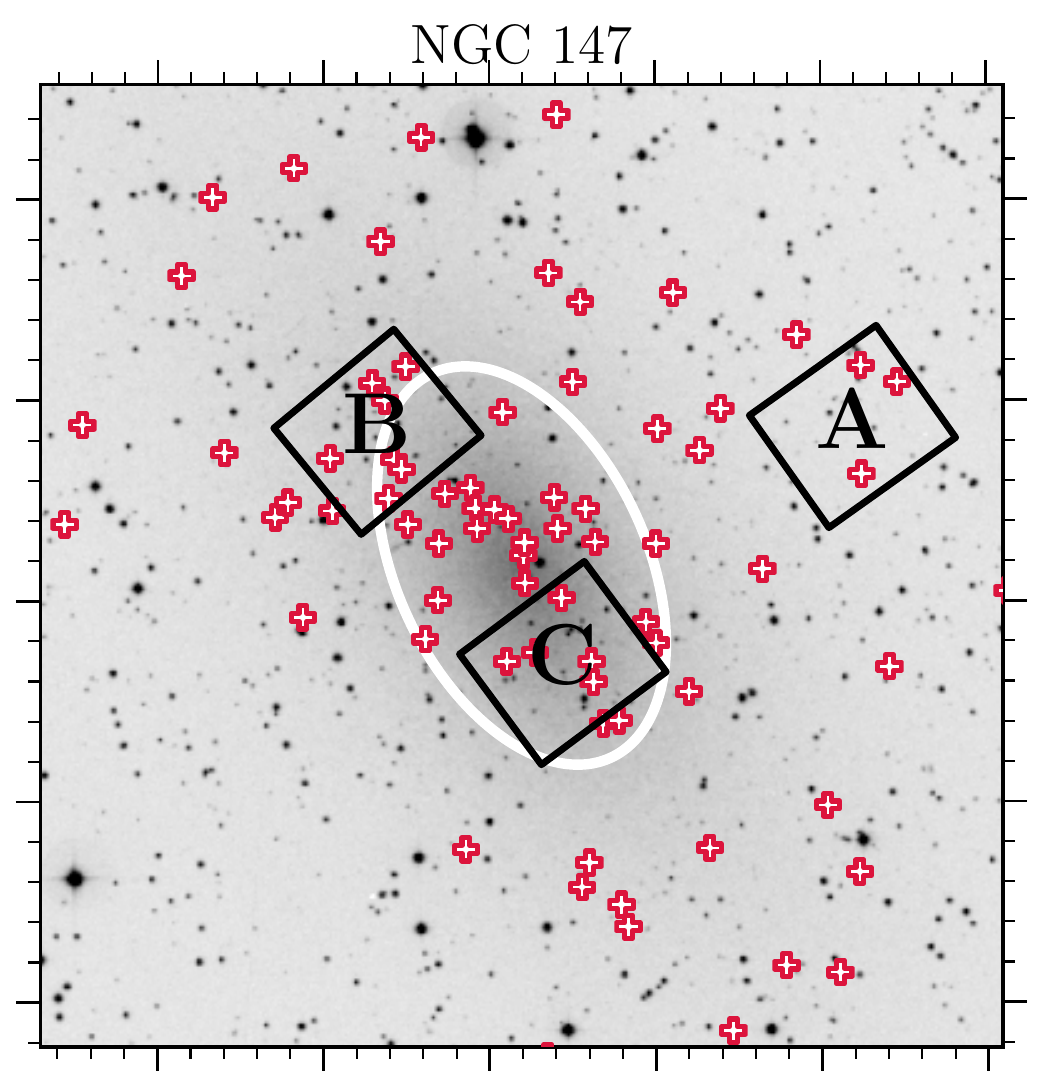}
\includegraphics[width=0.32\textwidth]{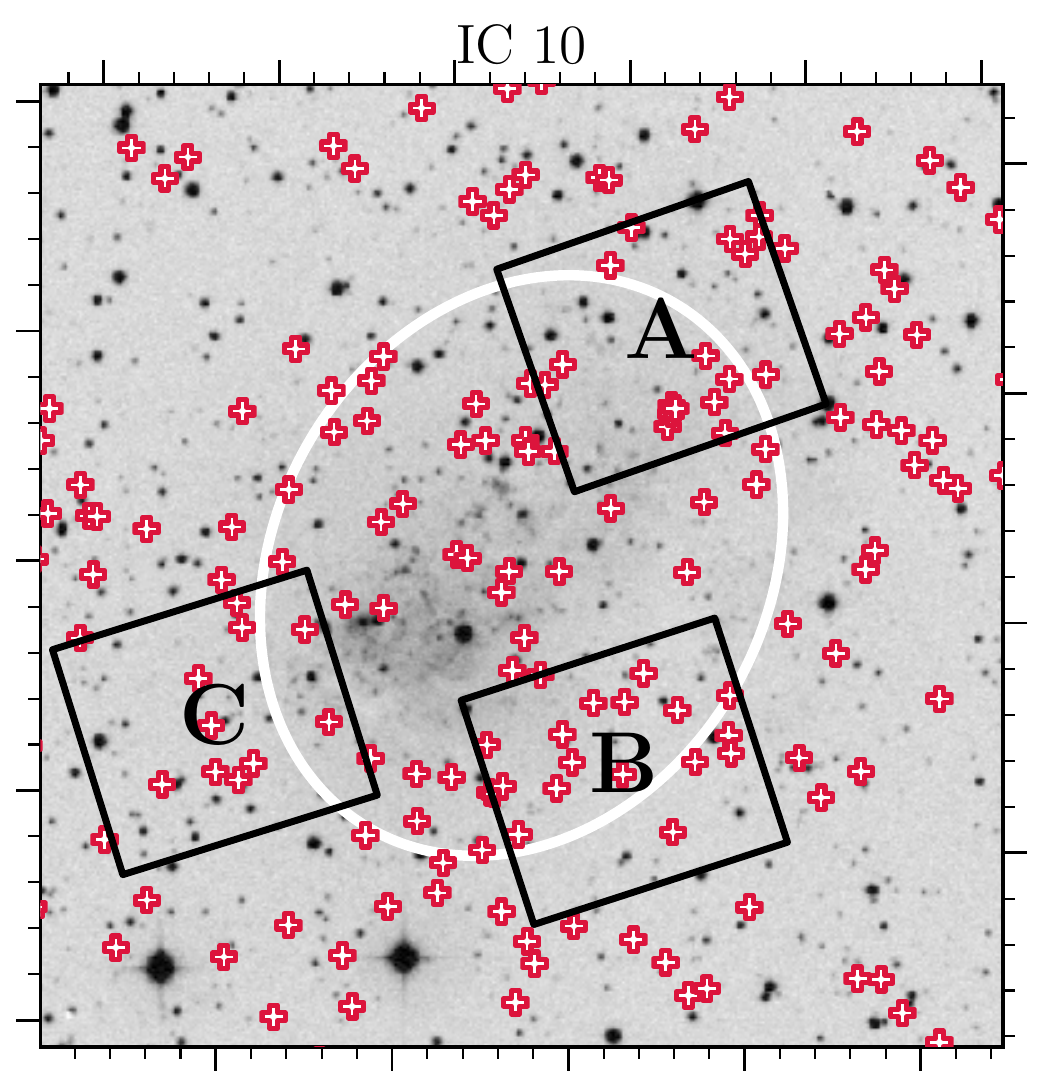}
\includegraphics[width=0.32\textwidth]{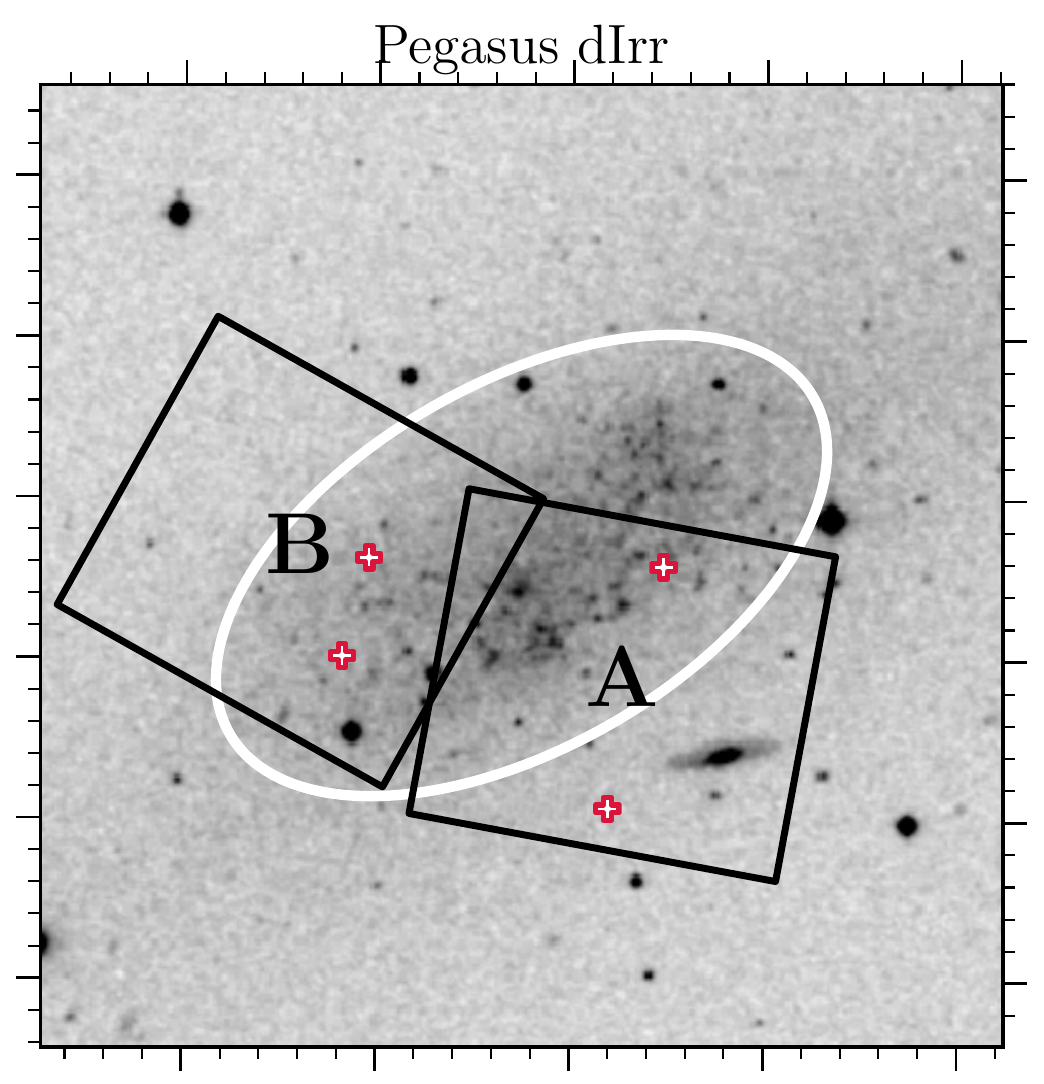}
}                                      
\hbox{                                 
\includegraphics[width=0.32\textwidth]{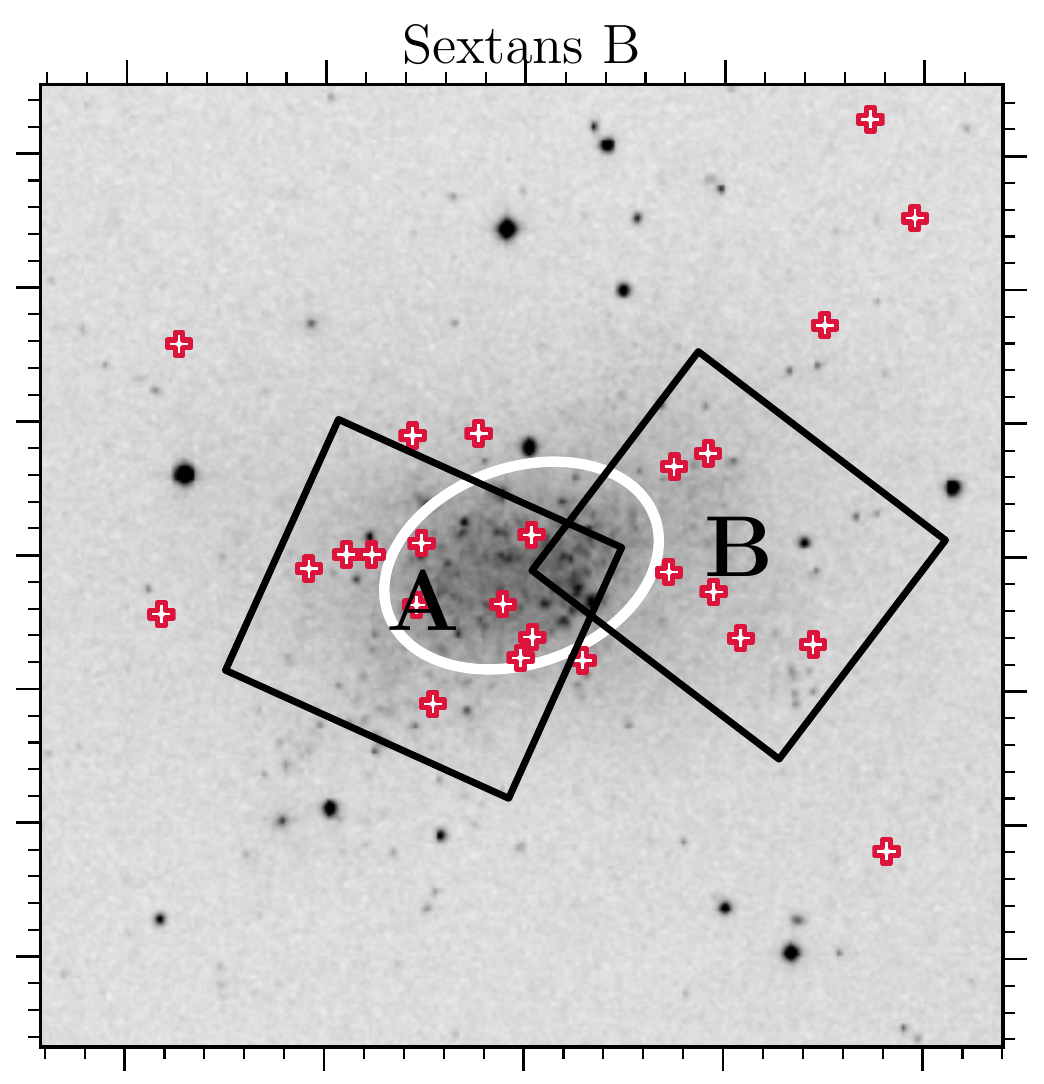}
\includegraphics[width=0.32\textwidth]{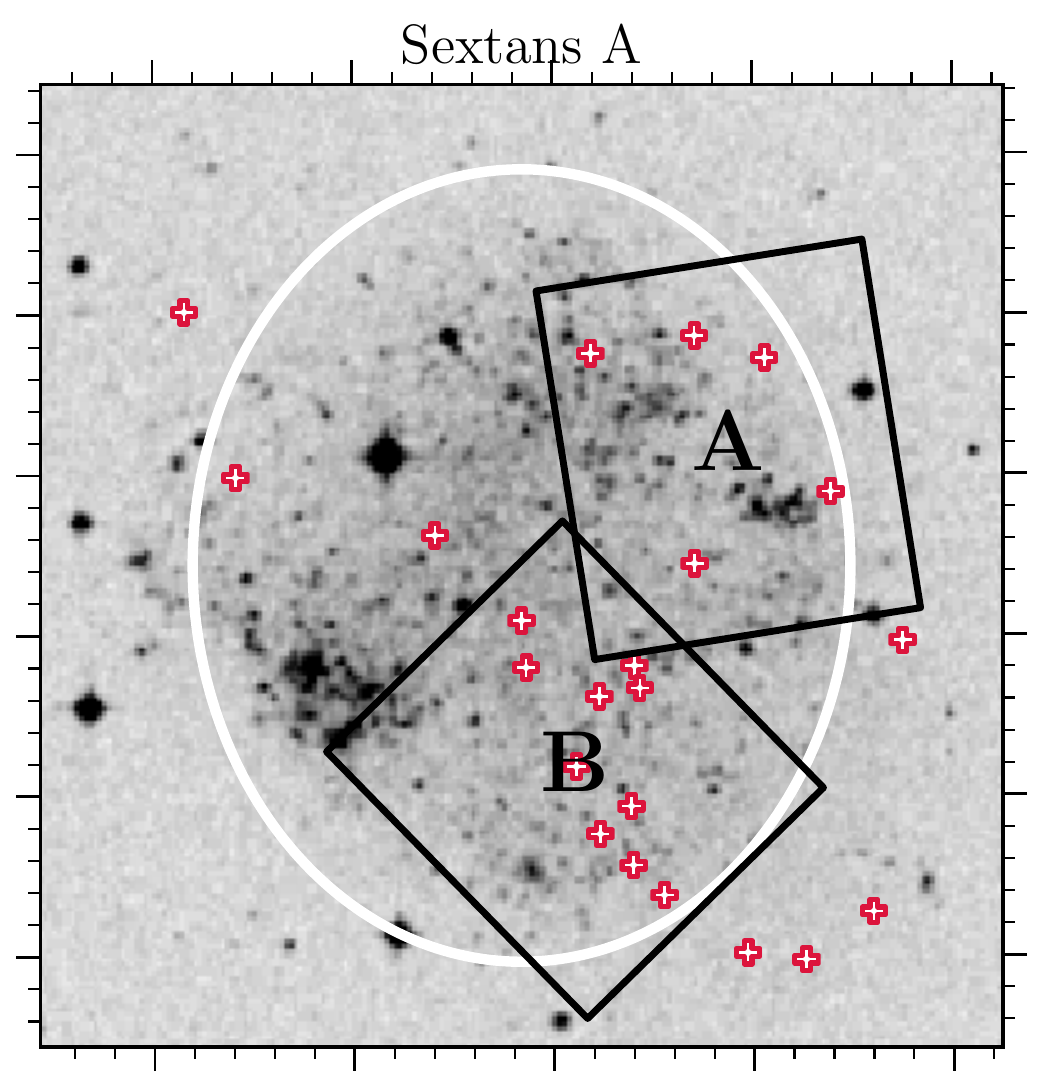}
\includegraphics[width=0.32\textwidth]{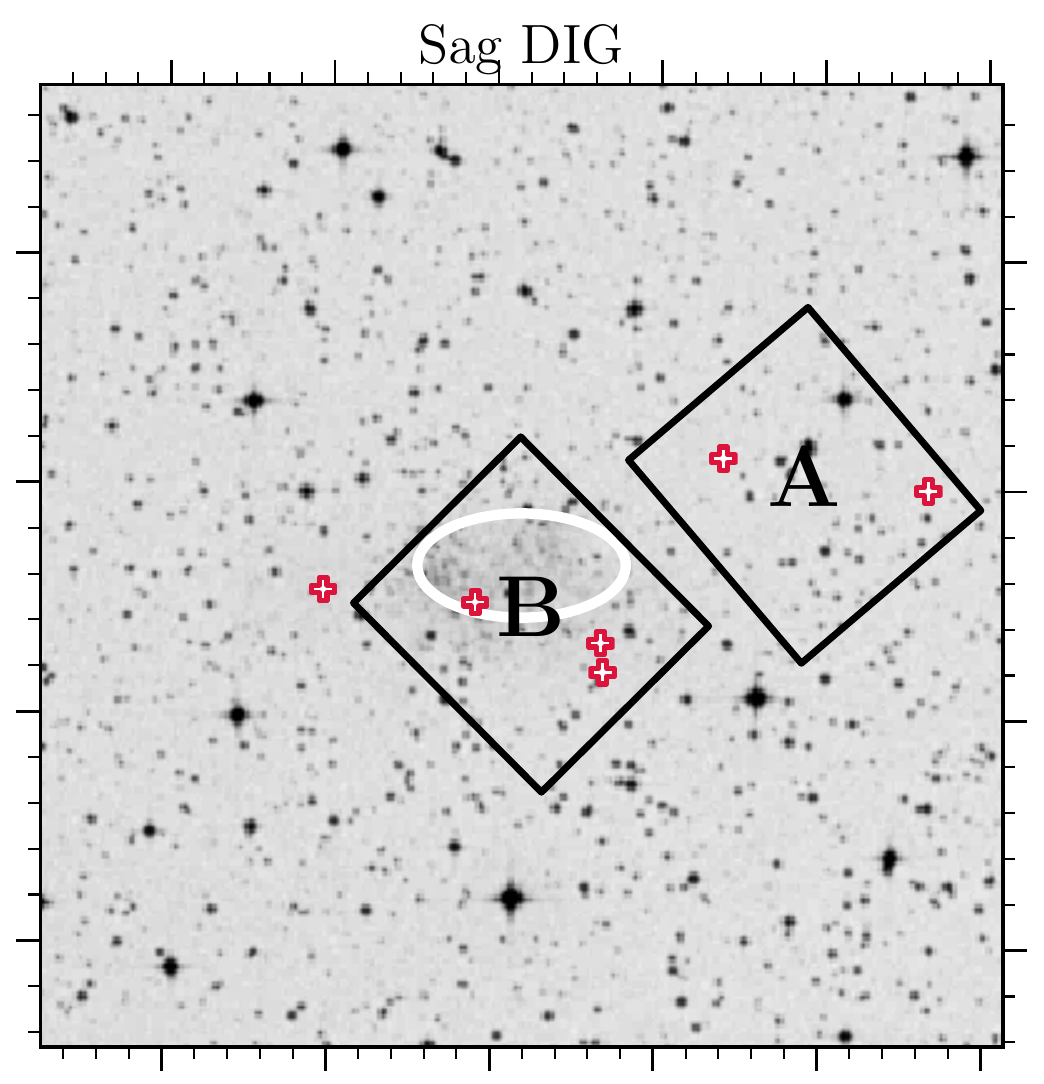}
}
}
\caption{DSS images of the 6 target galaxies with North pointing up
  and East to the left. Black boxes mark the HST footprints
  (123\arcsec $\times$ 136\arcsec), labeled A, B, and C
  (Table~\ref{tab:obs}). White ellipses mark the half-light radii of
  the galaxies. The dusty AGB variables from \dustings\ are marked
  with red plus symbols \citep{Boyer+2015b}. Fields were selected to
  maximize the number of \dustings\ variables covered and to include
  the dustiest examples. \label{fig:maps}}
\end{figure*}

\subsection{Sample Selection}
\label{sec:samp}

We observed six \dustings\ galaxies with {\it HST}'s WFC3/IR in Cycle
23 as part of GO-14073 from Oct 2016 to Aug
2016. Table~\ref{tab:targets} lists the observed galaxies and their
properties. We chose these six galaxies because they have large
populations of dust-producing AGB candidates and span the entire
metallicity range of the \dustings\ sample. Four of the galaxies are
gas-rich dwarf irregular (dIrr) systems with \ion{H}{2} regions that
point to sites of current massive star formation. One galaxy
(Pegasus/DDO\,216) is a transition-type galaxy (dTrans). It is
gas-rich, but there are no \ion{H}{2} regions. One galaxy (NGC\,147)
is a dwarf spheroidal (dSph). It is a gas-poor galaxy with no current
star formation, but a sizeable intermediate-aged AGB population is evident
\citep[e.g.,][]{Lorenz+2011,Golshan+2017}. The star-formation histories
of all six galaxies are described by \citet{Weisz+2014}.

\begin{deluxetable*}{rcccccccr}
\tablewidth{0pc}
\tabletypesize{\normalsize}
\tablecolumns{9}
\tablecaption{Observations\label{tab:obs}}
\tablehead{
\colhead{}&
\colhead{}&
\colhead{}&
\colhead{}&
\colhead{}&
\colhead{F127M}&
\colhead{F139M}&
\colhead{F153M}&
\colhead{}\\
\colhead{Galaxy}&
\colhead{Field}&
\colhead{R.A.}&
\colhead{Decl.}&
\colhead{Start Date}&
\colhead{$t_{\rm exp}$}&
\colhead{$t_{\rm exp}$}&
\colhead{$t_{\rm exp}$}&
\colhead{Orient.}\\
\colhead{}&
\colhead{}&
\colhead{(J2000)}&
\colhead{(J2000)}&
\colhead{(UT)}&
\colhead{(s)}&
\colhead{(s)}&
\colhead{(s)}&
\colhead{(E of N)}
}
\startdata
IC 10 & A & 00h20m08.22s & $+$59d20m25.68s & 2015-10-18 04:56:34 & 875.6 & 896.9 & 896.9 & 22\fdg9 \\
IC 10 & B & 00h20m08.83s & $+$59d16m37.24s & 2015-10-19 04:48:44 & 875.6 & 896.9 & 896.9 & 21\fdg6 \\
IC 10 & C & 00h20m36.95s & $+$59d16m48.82s & 2015-10-19 16:38:31 & 875.6 & 896.9 & 896.9 & 21\fdg1 \\
NGC 147 & A & 00h32m41.89s & $+$48d32m44.21s & 2015-10-09 22:28:15 & 846.9 & 875.6 & 846.9 & 35\fdg2 \\
NGC 147 & B & 00h33m24.92s & $+$48d32m40.26s & 2015-10-07 02:57:33 & 846.9 & 875.6 & 846.9 & 39\fdg5 \\
NGC 147 & C & 00h33m08.14s & $+$48d29m12.66s & 2015-10-09 02:59:39 & 846.9 & 875.6 & 846.9 & 36\fdg5 \\
Pegasus dIrr & A & 23h28m33.15s & $+$14d43m54.62s & 2015-10-02 06:57:09 & 796.9 & 825.6 & 796.9 & $-$10\fdg1 \\
Pegasus dIrr & B & 23h28m41.42s & $+$14d44m41.16s & 2015-10-04 07:07:30 & 796.9 & 825.6 & 796.9 & $-$28\fdg9 \\
Sag DIG & A & 19h29m48.07s & $-$17d39m59.21s & 2016-08-07 13:47:10 & 796.9 & 846.9 & 796.9 & $-$49\fdg0 \\
Sag DIG & B & 19h29m58.04s & $-$17d41m07.73s & 2016-08-07 14:46:50 & 796.9 & 846.9 & 796.9 & $-$44\fdg6 \\
Sextans A & A & 10h10m55.20s & $-$04d40m56.64s & 2016-03-06 04:03:59 & 796.9 & 825.6 & 796.9 & $-$81\fdg1 \\
Sextans A & B & 10h10m58.91s & $-$04d42m51.65s & 2016-04-04 01:21:55 & 796.9 & 825.6 & 796.9 & $-$45\fdg8 \\
Sextans B & A & 10h00m02.49s & $+$05d19m38.36s & 2016-03-14 02:54:20 & 796.9 & 825.6 & 796.9 & $-$24\fdg1 \\
Sextans B & B & 09h59m53.00s & $+$05d20m00.84s & 2016-03-17 17:08:40 & 796.9 & 825.6 & 796.9 & $-$37\fdg2\enddata

\tablecomments{\ This data is from \hst\ program GO-14073 (P.I. Boyer; doi:10.17909/T9HM3M).}
\end{deluxetable*}

The $2\farcm1 \times 2\farcm3$ WFC3/IR field of view is very small
compared to the \dustings\ footprint ($\gtrsim 10\arcmin \times
10\arcmin$), so we observed galaxies with 2 or 3 fields, placed to
maximize the coverage of \dustings\ AGB candidates while also
maximizing the inclusion of those with $[3.6]-[4.5] > 0.5$~mag, i.e.,
the dustiest stars. Figure~\ref{fig:maps} shows the placement of the
WFC3/IR fields. In total, we have covered 99 of the 375 original
\dustings\ x-AGB variables reported for these 6 galaxies in Paper~II.

\subsection{Observations \& Photometry}
\label{sec:obs}

Observations are summarized in Table~\ref{tab:obs}. We imaged each
field with the F127M, F139M, and F153M filters, employing four dithers
with the {\asciifamily WFC3-IR-DITHER-BOX-MIN} pattern to minimize
image defects and to maximize the spatial resolution (the resulting
mosaics are Nyquist sampled). The total exposure times are
796.9--896.9~s in each filter, depending on the detector sampling
sequence used.  WFC3/IR is non-destructively read out multiple times
during an exposure using sequences combining long and short reads that
provide uniform sampling over a wide range of stellar magnitudes
({\asciifamily MULTIACCUM}
mode\footnote{http://www.stsci.edu/hst/wfc3/documents/handbooks/\\currentIHB/c07\_ir08.html}). For
the first three exposures in each filter, we adopted sample sequence
{\asciifamily STEP100} with {\asciifamily NSAMP=8}. The fourth and
final exposure for each filter was set to fit within the remaining
orbit visibility, which differed for each galaxy: {\asciifamily
  STEP100,NSAMP=8}; {\asciifamily STEP100,NSAMP=9}; {\asciifamily
  STEP50,NSAMP=10}; or {\asciifamily SPARS25,NSAMP=12}.

We used \hst\ Drizzlepac v2.0 to create mosaics in each filter. We
combined the calibrated, flat-fielded exposures ({\asciifamily
  flt.fits} files) using {\sc Astrodrizzle} to create a stacked,
drizzled image with pixel scale 0\farcs0642/pix. Stellar positions
were measured on the drizzled F127M image, allowing for forced
photometry of faint stars in the individual images.

\begin{deluxetable}{ll}
\tablewidth{0pt}
\tablecolumns{2}
\tablecaption{Catalog Information\label{tab:cat}}
\tablehead{
\colhead{Column}&
\colhead{Column}\\
\colhead{Number}&
\colhead{Description}
}
\startdata
1 & Identification number \\
2 & \dustings\ ID\\
3 & Galaxy name\\
4 & Right Ascension (J2000)\\
5 & Declination (J2000) \\
6--7 & F127M magnitude and 1-$\sigma$ error\\
8--9 & F139M magnitude and 1-$\sigma$ error\\
10--11 & F153M magnitude and 1-$\sigma$ error\\
12--13 & [3.6] magnitude and 1-$\sigma$ error\tablenotemark{a}\\
14--15 & [4.5] magnitude and 1-$\sigma$ error\tablenotemark{a}\\
16 & C or M classification\\
17 & x-AGB variable Flag\tablenotemark{b}\\
18 & Dusty Flag\tablenotemark{c}\\
19 & Contamination Flag\tablenotemark{d}
\enddata

\tablecomments{\ The catalog is available with
  the electronic version of this paper and on VizieR.}

\tablenotetext{a}{\ Magnitudes are from the coadded epochs ($\approx$180~days separation). See Paper~I.}

\tablenotetext{b}{\ The x-AGB variables identified in Paper II.}

\tablenotetext{c}{\ Sources with $>$4-$\sigma$ excess in $[3.6]-[4.5]$.}

\tablenotetext{d}{\ Sources that are suspected contaminants
  (\S\ref{sec:cont}).}
\end{deluxetable}

\begin{figure}
  \includegraphics[width=\columnwidth]{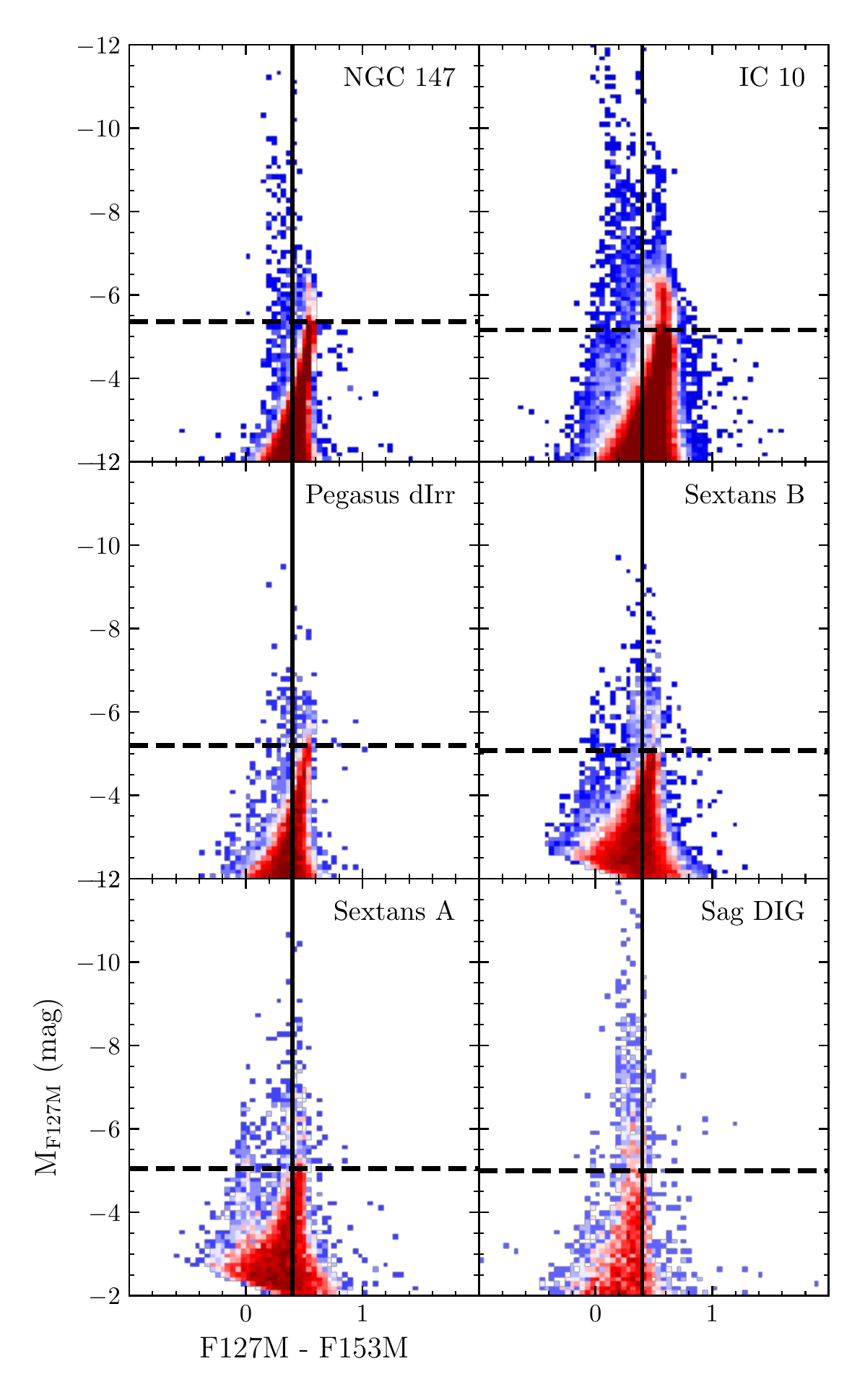}
   \caption{{\it HST} color-magnitude diagrams including stars with
     $S/N > 4$ (red: higher source density, blue: lower source
     density). The dashed red line marks the TRGB
     (\S\ref{sec:trgb}). The solid line at {\rm F127M-F153M = 0.4~mag}
     is for reference. The main sequence is visible in most panels at
     a color of 0~mag. Foreground stars are in the vertical sequence
     just to the left of the solid line and are especially prominent in
     IC\,10 and Sag\,DIG, which are near the Galactic
     Plane. \label{fig:cmds}}
\end{figure}

We performed PSF photometry on the {\asciifamily flt.fits} images
using {\asciifamily DOLPHOT's} WFC3-specific module
\citep{Dolphin2000}. We retain only stars with error flag $<$8 and
signal-to-noise $>$4 and require low sharpness and crowding values.
Restricting the sharpness parameter minimizes contamination from
extended sources and cosmic rays, and restricting the crowding
parameter ensures that a star's flux measurement is not substantially
affected by nearby stars. We adopt $(\Sigma\, {\rm Sharp}_\lambda)^2 <
0.1$ and $(\Sigma\, {\rm Crowd}_\lambda) < 1.5$~mag. The crowd
parameter is a measure of how much brighter a star would be if nearby
stars were not subtracted simultaneously. The sharp parameter is
positive for sharp sources (cosmic rays) and negative for extended
sources.  These initial quality cuts eliminate most contamination and
poor measurements from the catalog, but inspection of the images
suggests that some extended and/or blended sources remain. We
therefore flag sources with $(\Sigma\, {\rm Sharp}_\lambda)^2 > 0.02$
as those that require further scrutiny.

Artificial star tests indicate that photometry is $>$90\% complete at
$S/N > 10$. Since our AGB analysis is restricted to sources brighter
than this limit, our AGB samples are near-complete.

To identify dust-producing stars, we use \spitzer\ photometry from
Paper~I. We use the magnitudes measured after coadding the two
observing epochs to mitigate the effect of pulsation in the IR data.

The \hst\ catalog was matched to the \spitzer\ catalog using the
{\asciifamily DAOMaster} routine \citep{Stetson1987}, which
iteratively solves for the transformation coefficients and assigns
matches, starting with a 3\arcsec\ radius and decreasing to a 0\farcs6
radius. The \hst\ and \spitzer\ data are poorly matched in both
spatial resolution and sensitivity; to minimize mismatches between
these two datasets, we restrict the \hst\ input catalog to stars with
signal-to-noise ($S/N$) $>$10 in all three filters thereby excluding
the faintest stars that are unlikely to have
\spitzer\ counterparts. IC\,10 is the most densely populated galaxy in
our sample, and the lower-resolution \spitzer\ data are therefore
strongly affected by crowding (Paper~I). We thus restrict the IC\,10
\hst\ catalog to $S/N > 15$ to achieve a good match to the
\spitzer\ sources.  We manually checked all \dustings\ x-AGB variables
from Paper~II that were not matched to HST sources using these
criteria. Six \dustings\ variables match HST sources with
low $S/N$, and these were added back into the catalogs.

All photometry presented in this paper is corrected for extinction
using the $A_{\rm V}$ values listed by \citet{Weisz+2014}. That work
takes values from \citet{SchlaflyFinkbeiner2011} if $A_{\rm V} <
0.2$. Higher extinction values were estimated by comparing observed
and simulated optical color-magnitude diagrams. We assume $A_{\rm
  F127M}/A_{\rm V} = 0.27391$, $A_{\rm F139M}/A_{\rm V}=0.23979$,
$A_{\rm F153M}/A_{\rm V}=0.20366$, $A_{3.6}/A_{\rm V}=0.06706$, and
$A_{4.5}/A_{\rm V}=0.05591$, from the Padova
simulations.\footnote{http://stev.oapd.inaf.it/cgi-bin/cmd}

Figure~\ref{fig:cmds} shows the HST color-magnitude diagrams, and
Table~\ref{tab:cat} lists the columns included in the final catalog,
available to download in the electronic version of this paper and on
VizieR.  We have made no corrections to the $HST$ astrometry, which
has $\approx$200~mas absolute accuracy and $\approx$10~mas relative
accuracy.\footnote{http://www.stsci.edu/hst/wfc3/documents/handbooks/\\currentDHB/wfc3\_dhb.pdf}
The \spitzer\ positions are aligned to 2MASS by the \spitzer\
pipeline and thus have an absolute astrometric accuracy
$\approx$150~mas.

\subsection{Classification}

To select C- and M-type stars, we first measure the tip of the red
giant branch (TRGB).  Once sub-TRGB stars are eliminated, we use the
\citet{Aringer+2009,Aringer+2016} models to define color cuts that
separate AGB spectral types.  In all six galaxies, we find a total of
908 C stars and 2120 M stars; Table~\ref{tab:data} lists the final
adopted star counts.

\begin{deluxetable*}{lc|cccc|cccc}
\tablewidth{4in}
\tabletypesize{\normalsize}
\tablecolumns{10}
\tablecaption{Source Counts \label{tab:data}}
\tablehead{
\colhead{}&
\colhead{}&
\multicolumn{4}{c}{\rule[0.5em]{6em}{0.1pt}\ C\ \rule[0.5em]{6em}{0.1pt}}&
\multicolumn{4}{c}{\rule[0.5em]{6em}{0.1pt}\ M\ \rule[0.5em]{6em}{0.1pt}}
\\
\colhead{Galaxy}&
\colhead{$N_{\rm TRGB}$}&
\colhead{All}&
\colhead{Dusty\tablenotemark{a}}&
\colhead{Paper II\tablenotemark{b}}&
\colhead{Cont.\tablenotemark{c}}&
\colhead{All}&
\colhead{Dusty\tablenotemark{a}}&
\colhead{Paper II\tablenotemark{b}}&
\colhead{Cont.\tablenotemark{c}}
}
\startdata
IC 10 & 2928 & 531 & 49 & 26 & 5 & 1766 & 15 & 6 & 2 \\
NGC 147 & 370 & 65 & 21 & 14 & 1 & 265 & 2 & 1 & 3 \\
Pegasus & 154 & 44 & 8 & 4 & 1 & 32 & 0 & 0 & 3 \\
Sag DIG & 149 & 16 & 4 & 3 & 6 & 3 & 2 & 0 & 3 \\
Sextans A & 386 & 65 & 14 & 10 & 3 & 9 & 3 & 1 & 0 \\
Sextans B & 545 & 187 & 24 & 11 & 5 & 45 & 4 & 1 & 0\enddata
\tablecomments{\ $N_{\rm TRGB}$ indicates the number of stars brighter
  than the TRGB in F153M.  C- and M-type stars are identified by their
  HST near-IR colors (\S\ref{sec:cmclass}).}
\tablenotetext{a}{\ The subset of C and M stars with $[3.6]-[4.5]$
  color $>$4$\sigma$ that are not classified as contaminants. (\S\ref{sec:dustclass}).}
\tablenotetext{b}{\ The subset of C and M stars identified as x-AGB variables in Paper~II not classified as contaminants.}
\tablenotetext{c}{\ The subset of C and M stars that show $>$4$\sigma$ dust excess, but are suspected contaminants based on their location in Figure~\ref{fig:hcmd}.}

\end{deluxetable*}

\begin{deluxetable*}{rccccc}
\tablewidth{0pt}
\tabletypesize{\normalsize}
\tablecolumns{6}
\tablecaption{Near-IR TRGB\label{tab:trgb}}
\tablehead{
\colhead{}&
\multicolumn{4}{c}{TRGB}&
\colhead{}\\
\colhead{Galaxy}&
\colhead{F814W}&
\colhead{F127M}&
\colhead{F139M}&
\colhead{F153M}&
\colhead{$(m-M)_0$}\\
\colhead{}&
\colhead{(mag)}&
\colhead{(mag)}&
\colhead{(mag)}&
\colhead{(mag)}&
\colhead{(mag)}
}
\startdata
N\,147 &24.39$\pm$0.06& 19.03$\pm$0.04 & 18.82$\pm$0.04 & 18.45$\pm$0.04 & 24.39$\pm$0.06\\
IC\,10 &24.43$\pm$0.03& 19.27$\pm$0.04 & 19.06$\pm$0.04 & 18.68$\pm$0.04 & 24.43$\pm$0.03\\
Peg    &24.96$\pm$0.04& 19.76$\pm$0.05 & 19.55$\pm$0.05 & 19.20$\pm$0.05 & 24.96$\pm$0.04\\
Sex\,B &25.77$\pm$0.03& 20.71$\pm$0.05 & 20.52$\pm$0.04 & 20.23$\pm$0.05 & 25.77$\pm$0.03\\
Sex\,A &25.82$\pm$0.03& 20.77$\pm$0.09 & 20.58$\pm$0.09 & 20.32$\pm$0.09 & 25.82$\pm$0.03\\
Sag    &25.18$\pm$0.04& 20.18$\pm$0.06 & 20.03$\pm$0.06 & 19.77$\pm$0.07 & 25.18$\pm$0.04\enddata

\tablecomments{\ Distance modulii are derived from the F814W TRGB \citep{McQuinn+2017}, using the relationship from \citep{Rizzi+2007}. These are within 1\,$\sigma$ of distance modulii listed by \citet{McConnachie2012}, except Pegasus\,dIrr (2\,$\sigma$) and NGC\,147 (3\,$\sigma$).}

\end{deluxetable*}

\subsubsection{TRGB}
\label{sec:trgb}

To find the TRGB in each filter, we follow the commonly used strategy
described by \citet{Mendez+2002}. First, we select stars with ${\rm
  F129M-F153M}>0.1$~mag to eliminate main sequence stars. Next, we
pass the Gaussian-smoothed luminosity function through a Sobel
edge-detection filter. We perform 500 Monte Carlo Bootstrap resampling
trials, each time adding 4$\sigma$ Gaussian random photometric errors
and random variations on the bin size and starting magnitude of the
luminosity function. The final TRGB and its uncertainty are the mean
and standard deviation of a Gaussian function fit to the TRGB
estimates from the 500 trials (Table~\ref{tab:trgb}). Other techniques
to find the TRGB can result in more precise measurements than the
edge-detector method we employ, such as Bayesian maximum likelihood
\citep[e.g.,][]{Makarov+2006, Conn+2012, McQuinn+2016}. However, since
our red giant branches are deep and well populated, Bayesian techniques result in
only a marginal gain in accuracy and reliability. 

The TRGBs are fairly stable against metallicity in these filters,
similar to the F814W TRGB \citep[e.g.,][]{Rizzi+2007}. Using the
distance modulii listed in Table~\ref{tab:trgb}, all three filters
show brighter TRGBs at higher metallicity, with the F153M filter
showing the largest metallicity effect ($\Delta {\rm TRGB
  (F127M/F139M/F153M)} = 0.35/0.42/0.53$~mag).

\begin{figure}
  \includegraphics[width=\columnwidth]{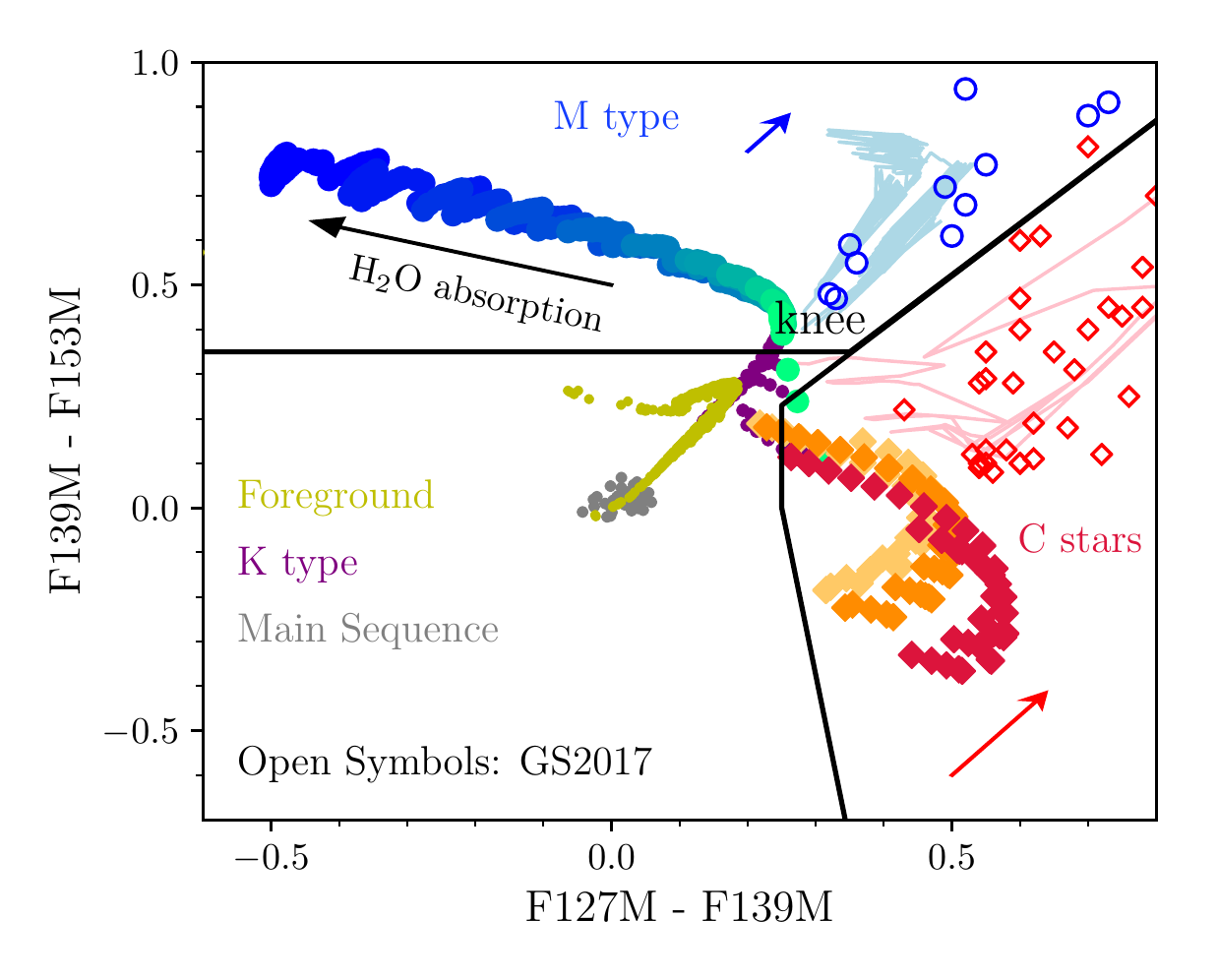}
  \caption{{\it HST} color-color diagram (CCD) showing the colors of C
    star models \citep[orange/red filled diamonds;][]{Aringer+2009}
    and M/K star models \citep[purple/cyan/blue filled
      circles;][]{Aringer+2016}. Only models brighter than the
      TRGB (horizontal line in Fig.~\ref{fig:hcmd}) are plotted. Also shown
    in yellow are simulated foreground star colors
    \citep{Girardi+2005} and a representative sample of main
    sequence stars in IC\,10 (gray; F127M-F153M < 0.1~mag). K stars
    are marked in dark purple.  The black lines mark the adopted C-
    and M-type regions. Red and blue arrows show the direction and
    magnitude of circumstellar extinction for C- and M-type stars,
    respectively (see text). The open circles/diamonds are dusty M/C
    stars in the LMC from \citet{GroenewegenSloan2017} and the
    light blue/pink lines show O-rich/C-rich COLIBRI models with
    \citet{Nanni+2016} dust growth models ($\log \epsilon_S =
    -13)$.\label{fig:modccd}}
  \end{figure}

\begin{figure*}
  \includegraphics[width=\textwidth]{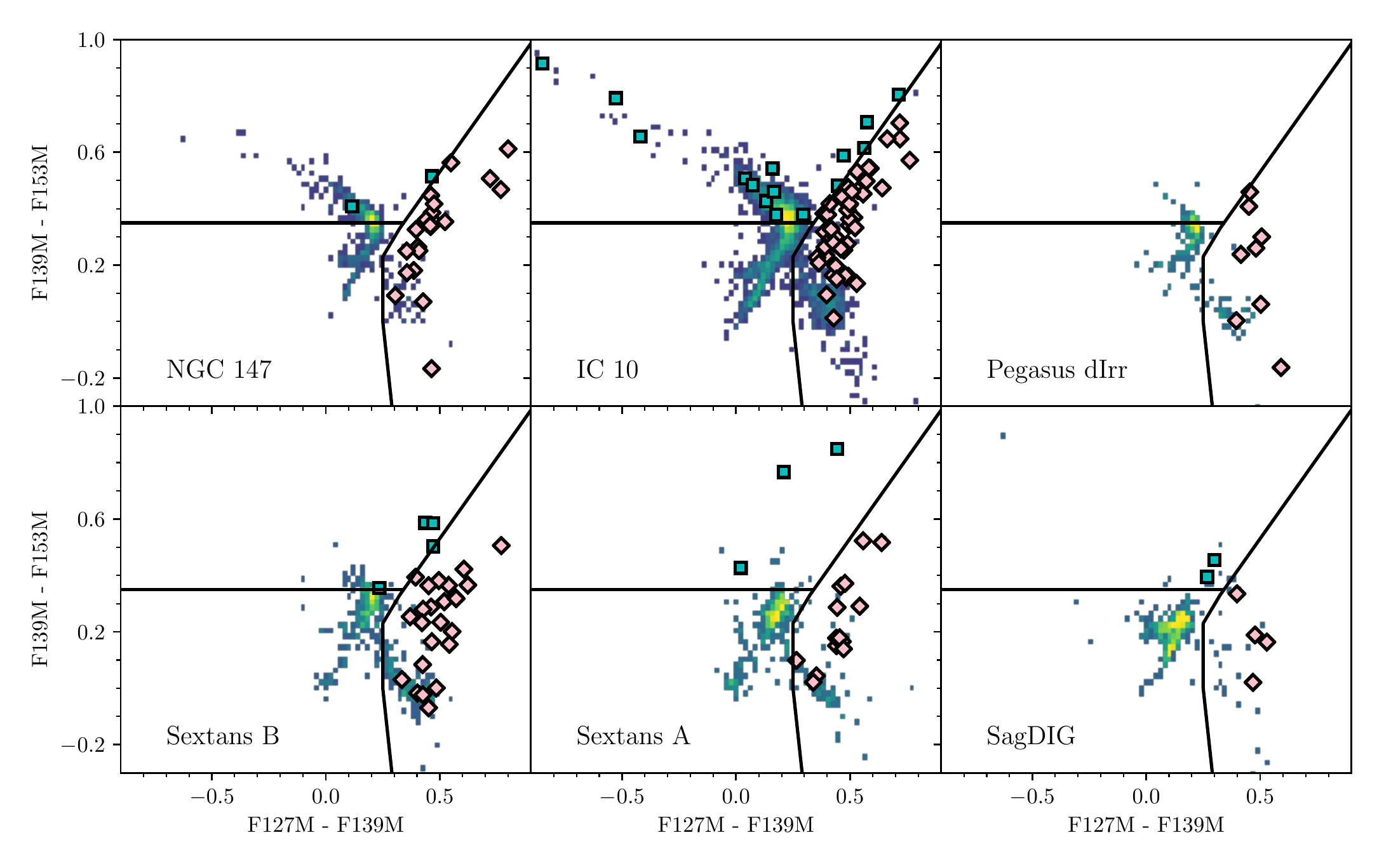}
   \caption{\hst\ color-color diagrams. Stars are included if they
     are brighter than the F153M TRGB. To retain the dustiest
     examples, stars fainter than this limit are also included if they
     are brighter than the 3.6~\micron\ TRGB and show $>$4$\sigma$
     excess in the $[3.6]-[4.5]$ color. Solid lines mark
     the adopted color cuts chosen to include only stars redder than
     the `knee' in F139M--F153M (M$0+$). The same cuts are used
     for all 6 galaxies.  Large cyan squares and pink diamonds mark
     the dusty M and C stars, respectively, identified by their {\it
       Spitzer} colors (see \S\ref{sec:dust}). \label{fig:ccds}}
\end{figure*}

\subsubsection{Color-Color Selection}
\label{sec:cmclass}

To select C- and M-type stars, we start with all sources brighter than
the F153M TRGB. This restriction results in the loss of the dustiest
AGB stars, which are made faint in F153M by circumstellar
extinction. We recover these stars in \S\ref{sec:dustclass}. 

Next, we classify the resulting subset of stars based on their
location in the F127M--F139M vs. F139M--F153M color-color diagram
(CCD; Fig.~\ref{fig:modccd}). When sub-TRGB stars are excluded (as
they are here), this diagram has three main branches: an M-type
branch, a C star branch, and a branch that includes a mixture of
K-type stars, main sequence stars, and foreground stars. We place the
divisions in Figure~\ref{fig:modccd} following the C, M and K star
models from \citet{Aringer+2009,Aringer+2016}. The C star models
included in Figure~\ref{fig:modccd} (plus symbols) are those with
$-0.8 \leq \log g\ {\rm [cm/s^2]} \leq 0$ and C/O = 1.4 (light
orange), 2 (orange), and 5 (red). For the M/K stars (crosses), we
include $T_{\rm eff} \leq 3700$~K and $-1 \leq \log g\ {\rm [cm/s^2]}
\leq -0.5$; darker blue colors represent lower effective
temperatures. Only models with ${\rm [Fe/H] \leq -1}$ are
included. The solid lines mark the adopted C- and M-type color
divisions. Red and blue arrows illustrate the direction and magnitude of
circumstellar extinction for $E(J-K_S)=1$~mag for
M-type stars with 60\% silicate + 40\% AlOx and C stars with 70\%
amorphous carbon + 30\% SiC \citep{Groenewegen2006}.

Both the C- and M-type models are slightly bluer than the data in both
colors. This discrepancy (approximately 0.05~mag) has no effect on the
C star definition since C stars are fairly well isolated from both M-
and K-type stars. However, the discrepancy is large enough to substantially
affect the number of M-type stars selected because of strong
contamination from warmer K-type stars. The reason for the slight
shifts between the models and data is unclear -- one possibility is a
bias in our adopted $A_{\rm V}$ values
(Table~\ref{tab:targets}). Another culprit might be the adopted water
opacity in the models from \citet{Aringer+2016}, who note that other
opacities shift the near-IR colors by about 0.05~mag. Further,
deviations from hydrostatic, spherically-symmetric atmospheres and
non-LTE conditions can affect the water opacity.  Because of this
slight color mismatch between the data and the models, we cannot use
the exact colors of the models to define the transition from K-type
stars to M-type stars (at around 3600~K) in the data.  Instead, we
note that M$0+$ models occupy the M/K-star sequence beginning
just blueward of the knee in the CCD at ${\rm
  F127M-F139M}\approx0.2$\,mag and ${\rm F139M-F153M}\approx0.4$\,mag.
We therefore select M-type stars based on the location of this knee in
the data (Fig.~\ref{fig:ccds}).

The adopted (F127M--F139M, F139M--F153M) positions of the C and M
divisions in the CCD are as follows: 

\begin{multline}
  {\rm M\ stars:}\ [(-0.6,0.35),(0.35,0.35),(1.0,1.1)]
\end{multline}

\begin{multline}
  {\rm C\ stars:}\\ [(0.35,-0.75),(0.25,0.00),(0.25,0.23),(1.0,1.1)]
\end{multline}

\noindent The line between C- and M-type stars on the right side of
the CCD follows the direction of circumstellar
extinction. \citet{GroenewegenSloan2017} (GS2017) fit $\sim$500 dusty
Magellanic Cloud stars using a modified version of the {\sc DUSTY}
radiative transfer code \citep{Groenewegen+2012}, and we passed the
resulting spectral energy distributions (SEDs) through the medium-band
HST filters used here. GS2017 used high-quality, multi-epoch data
  in the near-IR, so the SEDs used to derive the magnitudes in
  Figure~\ref{fig:modccd} are well constrained around 1~\micron. The
  C- or M-type classification in GS2017 is based on features in the
  stars' mid-IR spectra \citep[e.g.][]{Jones+2017}.  We include the
dustiest LMC stars from the GS2017 sample in Figure~\ref{fig:modccd},
with C-type stars as red open diamonds and M-type stars as blue open
circles. With the exception of one star in Figure~\ref{fig:modccd},
the dusty C- and M-type stars stay separated on the \hst\ CCD, indicating that contamination across the dividing line is low. For
  reference, we also show the COLIBRI tracks \citep{Marigo+2017}, with
  \citet{Nanni+2016} dust growth models.

We note that the GS2017 stars represent the dustiest examples of
  AGB stars. Less dusty examples that have not completely veiled their
  molecular features can fall along the entire H$_2$0 and C/O
  sequences, with dust affecting the color as indicated by the blue
  and red extinction arrows in Figure~\ref{fig:modccd}.

\subsubsection{Dusty Stars}
\label{sec:dustclass}

\begin{figure}
  \includegraphics[width=\columnwidth]{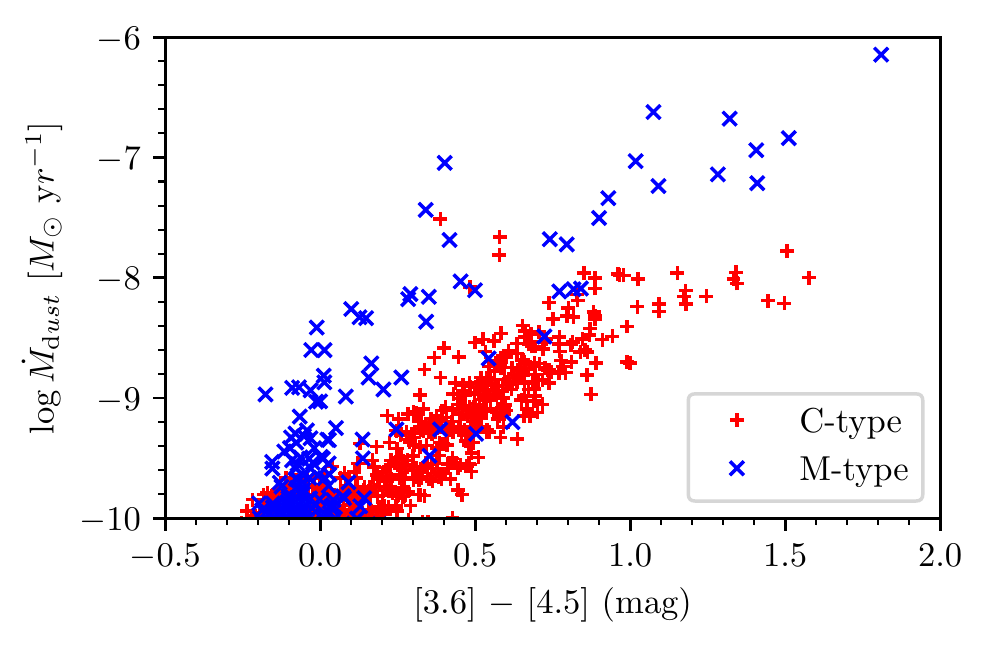}
  \caption{The relationship between dust-production rate and
    $[3.6]-[4.5]$ color for SMC AGB stars \citep{Srinivasan+2016}. The
    dust-production rate increases with color for $[3.6]-[4.5] \gtrsim
    0.1$~mag.\label{fig:grams}}
\end{figure}

Dusty AGB stars can be fainter than the TRGB both due to variability
and to circumstellar extinction. To recover these sources, we turn to
the \spitzer\ data. For both O-rich and C-rich stars in the Magellanic
Clouds, the $[3.6]-[4.5]$ color is approximately proportional to the
dust-production rates \citep{Riebel+2012, Srinivasan+2016,
  Sloan+2016}, especially at color $>$0.1~mag
(Fig.~\ref{fig:grams}). We therefore use the $[3.6]-[4.5]$ color
  as a proxy for dust excess, which we measure by comparing a source's
  color to the mean color of stars within 1~mag bins at 4.5~\micron.
  We flag sources as dusty if the excess exceeds 4$\sigma$. Dusty
objects that are fainter than the F153M TRGB are included in our
sample if they are brighter than the 3.6~\micron\ TRGB (Paper
III). There are examples of dusty carbon stars that are fainter than
the 3.6~\micron\ TRGB \citep[e.g., in the LMC;][]{Gruendl+2008}, but
given the small stellar masses of our galaxies relative to the LMC, we
expect to miss only a few such stars, if any, which does not impact
our star counts. Once the dusty stars are recovered, we classify them
as C- or M-type using the criteria outlined in \S\ref{sec:cmclass}.


Paper~II adopted the definition set by \citet{Blum+2006}, who defined
the dustiest AGB stars (dubbed extreme AGB or x-AGB stars) in the LMC
as those with $J-[3.6]>3.1$~mag.  IR spectroscopy has confirmed that
x-AGB stars are dominated by carbon stars, though there are also
O-rich examples
\citep{Trams+1999,vanLoon+2008b,Ruffle+2015,Boyer+2015c,Jones+2017}. Most
of the variable stars detected in the \dustings\ survey occupy the
same space in $[3.6]$ versus $[3.6]-[4.5]$ as the Magellanic Cloud
x-AGB stars, and were classified as such. The HST coverage includes 99
of the \dustings\ x-AGB variables. Of these, 90 are included in our
HST catalog and 77 are confidently identified as C- or M-type
(Table~\ref{tab:data}). Seven did not fall within the C or M
regions of the HST CCD and six are identified as possible contaminants
(see \S\ref{sec:cont}).

\begin{figure}
  \includegraphics[width=\columnwidth]{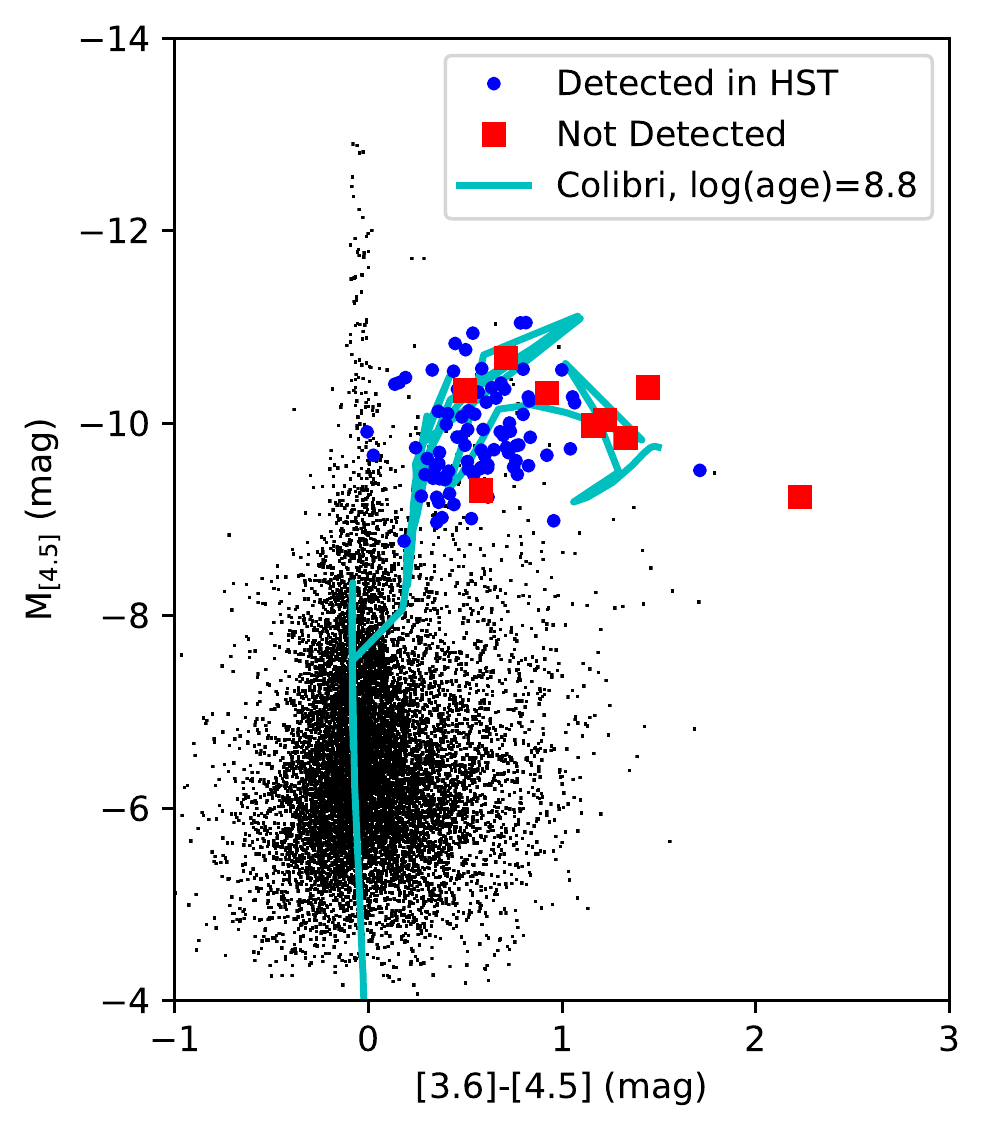}
  \caption{{\it Spitzer} color-magnitude diagram highlighting the 9
    \dustings\ sources not detected in the near-IR with HST. Black
    dots are all of the \dustings\ sources. Blue points are
    \dustings\ x-AGB variables with HST counterparts and red squares
    are those without HST counterparts.  A COLIBRI isochrone with
    $\log({\rm age}) = 8.8$ and \citet{Nanni+2016} dust growth models
    ($\log \epsilon_S = -13)$ is shown in cyan. The
    \dustings\ x-AGB variables (both those with and without HST
    counterparts) generally follow the isochrone.\label{fig:not}}
\end{figure}

Visual inspection verifies that nine \dustings\ x-AGB variables do not
have HST counterparts (Table~\ref{tab:nohst}). These nine sources have
an average IR color of $[3.6]-[4.5] = 1.1$~mag (Fig.~\ref{fig:not}),
and include 5 of the 6 dustiest sources identified in Paper II; those
with slightly bluer {\it Spitzer} colors in Figure~\ref{fig:not} may
be somewhat affected by CO absorption in the 4.5~\micron\ band.  All
nine sources are well isolated within $r \gtrsim 3$~pix in the HST
images, which eliminates crowding as a factor in their non-detection
and suggests that they are truly fainter than the sensitivity limit of
our HST observations. Their {\it Spitzer} colors closely follow
COLIBRI AGB tracks, and we therefore propose that these are AGB stars
with very strong circumstellar dust extinction.  In fact, their
  IR colors and absolute magnitudes are similar to the progenitor of
  the intermediate-luminosity optical transient SN2008S, which may
  have been an electron-capture SN from a massive AGB star
  \citep{Prieto+2008, Kahn+2010}. Alternatively, they could be other
red objects such as active galactic nuclei.

\begin{deluxetable}{lcccc}
\tablewidth{0pt}
\tabletypesize{\normalsize}
\tablecolumns{5}
\tablecaption{DUSTiNGS x-AGB Variables without HST Counterparts \label{tab:nohst}}
\tablehead{
  \colhead{Galaxy}&
  \colhead{ID}&
  \colhead{[3.6]}&
  \colhead{[4.5]}&
  \colhead{Amp.}\\
  \colhead{}&
  \colhead{}&
  \colhead{(mag)}&
  \colhead{(mag)}&
  \colhead{(mag)}
}
\startdata
IC 10 & 115785 & 17.42$\pm$0.06 & 15.20$\pm$0.03 & 0.62\\
IC 10 & 109882 & 15.61$\pm$0.03 & 14.39$\pm$0.03 & 0.45\\
IC 10 & 110204 & 15.91$\pm$0.05 & 14.58$\pm$0.03 & 0.85\\
IC 10 & 111624 & 15.71$\pm$0.07 & 15.13$\pm$0.03 & 0.74\\
Sex A &  90428 & 17.01$\pm$0.05 & 15.84$\pm$0.05 & 0.50\\
Sex A &  94477 & 16.89$\pm$0.05 & 15.45$\pm$0.03 & 0.67\\
Sex B &  85647 & 15.93$\pm$0.02 & 15.43$\pm$0.03 & 0.21\\
Sex B &  93730 & 16.38$\pm$0.03 & 15.46$\pm$0.03 & 0.40\\
Sex B &  96433 & 15.81$\pm$0.02 & 15.10$\pm$0.03 & 0.33
\enddata
\tablecomments{\ The ID is the \dustings\ ID from Papers I and II.  The amplitude is the change in 3.6~\micron\ magnitude between the 2 \dustings\ epochs (larger amplitudes are generally indicative of more dust).}
\end{deluxetable}

By combining the HST and \spitzer\ data, we find additional examples
of both M- and C-type dusty stars that were not identified as variable
x-AGB stars in Paper~II, most likely because they were observed at an
unfavorable pulsation phase. Altogether, we increase the number of
known dusty AGB stars by approximately 50\% (Table~\ref{tab:data}). Most of the x-AGB stars
that we identify as AGB stars here are C stars, similar to what is
seen in the Magellanic Clouds. However, there are examples of M-type
x-AGB stars, most notably in IC\,10 (see \S\ref{sec:mdust}).

\begin{figure*}
  \includegraphics[width=\textwidth]{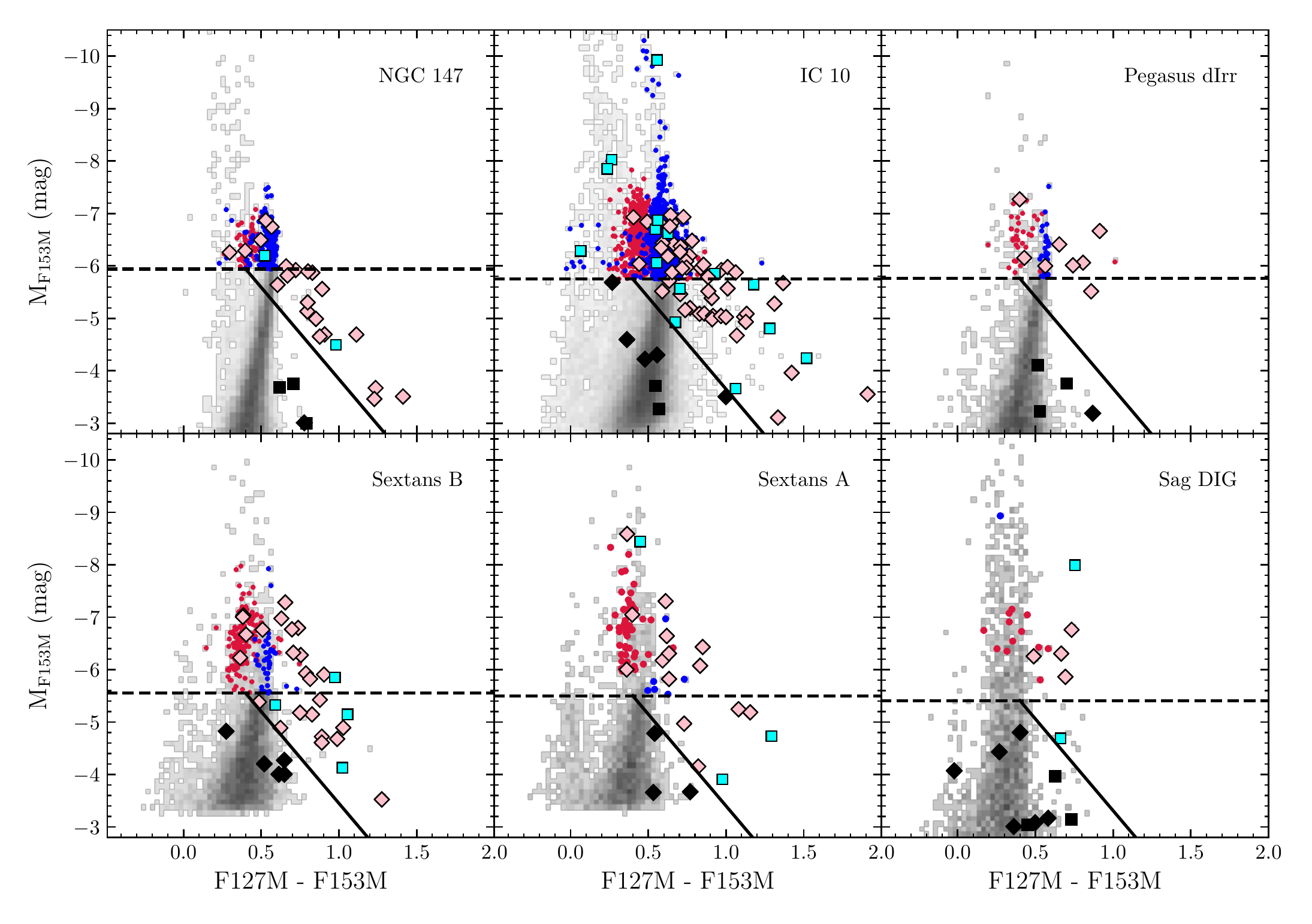}
   \caption{HST color-magnitude diagrams with C (red dots) and M (blue
     dots) stars labeled. Note,
     late-M-type stars can be very blue in F127M--F153M due to deep
     water absorption that affects F153M as well as
     F139M. Large cyan squares are M-type stars with
     $>$4$\sigma$ excess in the $[3.6]-[4.5]$ color, indicating the
     presence of circumstellar dust. Likewise, large pink diamonds are
     dusty C stars. The dashed line marks the TRGB and the solid line
     marks the adopted division between likely dusty AGB stars and
     contaminating sources (see text).  Contaminants (YSOs, PNe, etc.)
     are marked by solid black squares and diamonds, where the shape
     indicates whether they were classified as M or C stars,
     respectively, in the HST CCD (Fig.~\ref{fig:ccds}).  \label{fig:hcmd}}
\end{figure*}

\subsubsection{Contamination From K-type Giants}
\label{sec:cmcont}

Warmer stellar temperatures in metal-poor systems result in large
populations of K-type stars, and it is likely that there is some
contamination in both our C- and M-type samples.  The super-solar
metallicity galaxy M31 shows almost no contamination from K giants in
the same CCD used here \citep{Boyer+2013}, while Sextans\,A and
Sag\,DIG show a substantial population of K giants. This is evidenced
by the continuous sequence downwards from the knee in the CCD through
the entire M/K star model sequence and into the region of the CCD
dominated by foreground.

Figure~\ref{fig:modccd} shows a few K-type models within the
C star region. We have placed the boundaries of the C star
region to minimize this contamination, using the natural breaks in
source density between the M and C star population as a guide
(Fig.~\ref{fig:ccds}).

The M-type sample is highly susceptible to contamination from K-type
stars.  This is by far the largest source of contamination among the C
and M samples, and the strength of the contamination increases in
metal-poor galaxies.  Even a slight shift in the M/K star division has
a strong effect on the ratio of C- to M-type stars, and this is
discussed further in Paper~V (Boyer et al. 2017, in preparation). 

\subsection{Contamination From Other Objects}
\label{sec:cont}

We do not expect significant contamination from other source types
among our sample. Possible contaminants include young stellar objects
(YSOs), planetary nebulae (PNe), and post-AGB stars, but the
comparatively short lifetimes of these objects make them even more
rare than AGB stars.

\begin{figure*}
  \hbox{
    \includegraphics[width=\columnwidth]{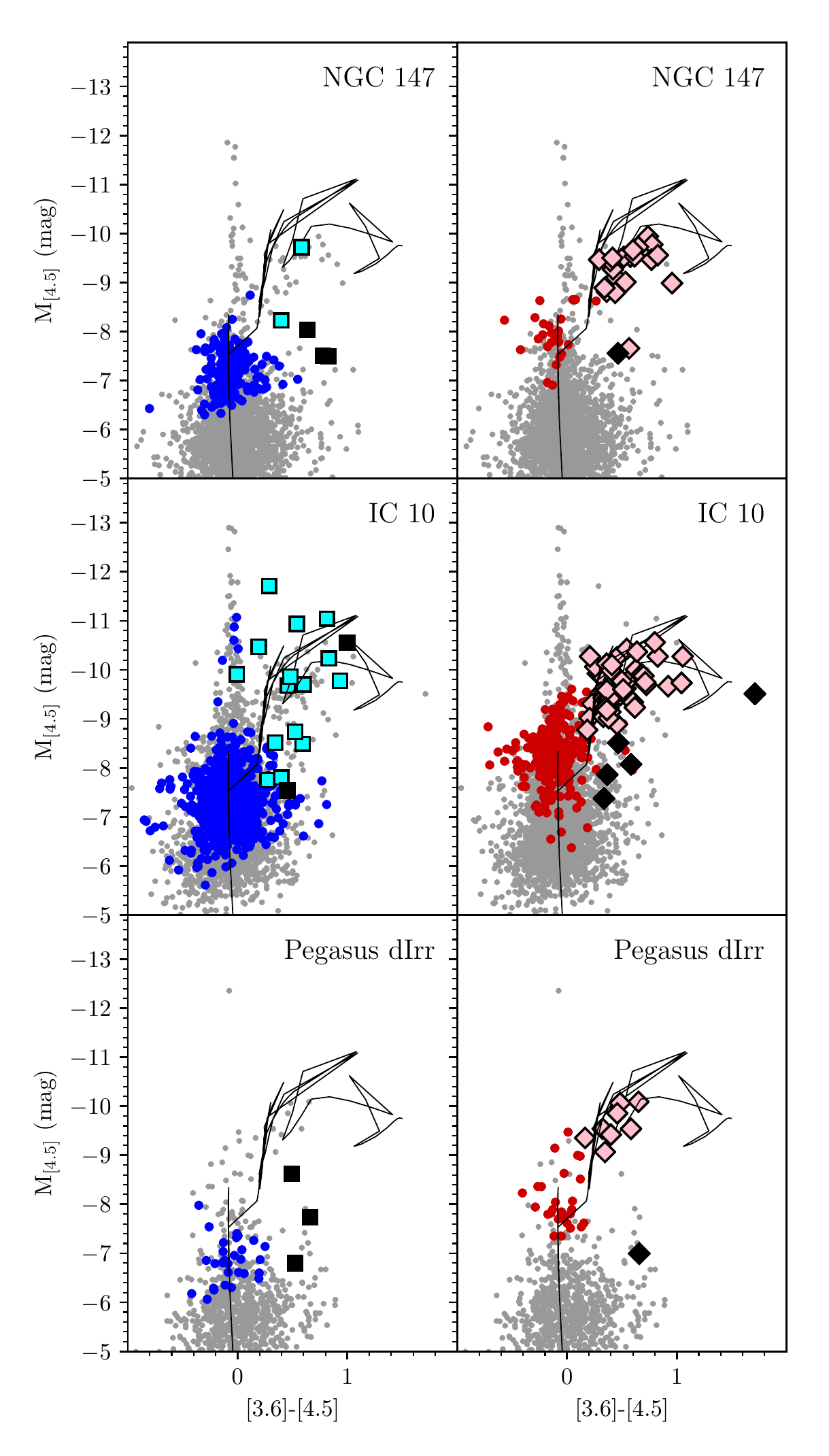}
    \includegraphics[width=\columnwidth]{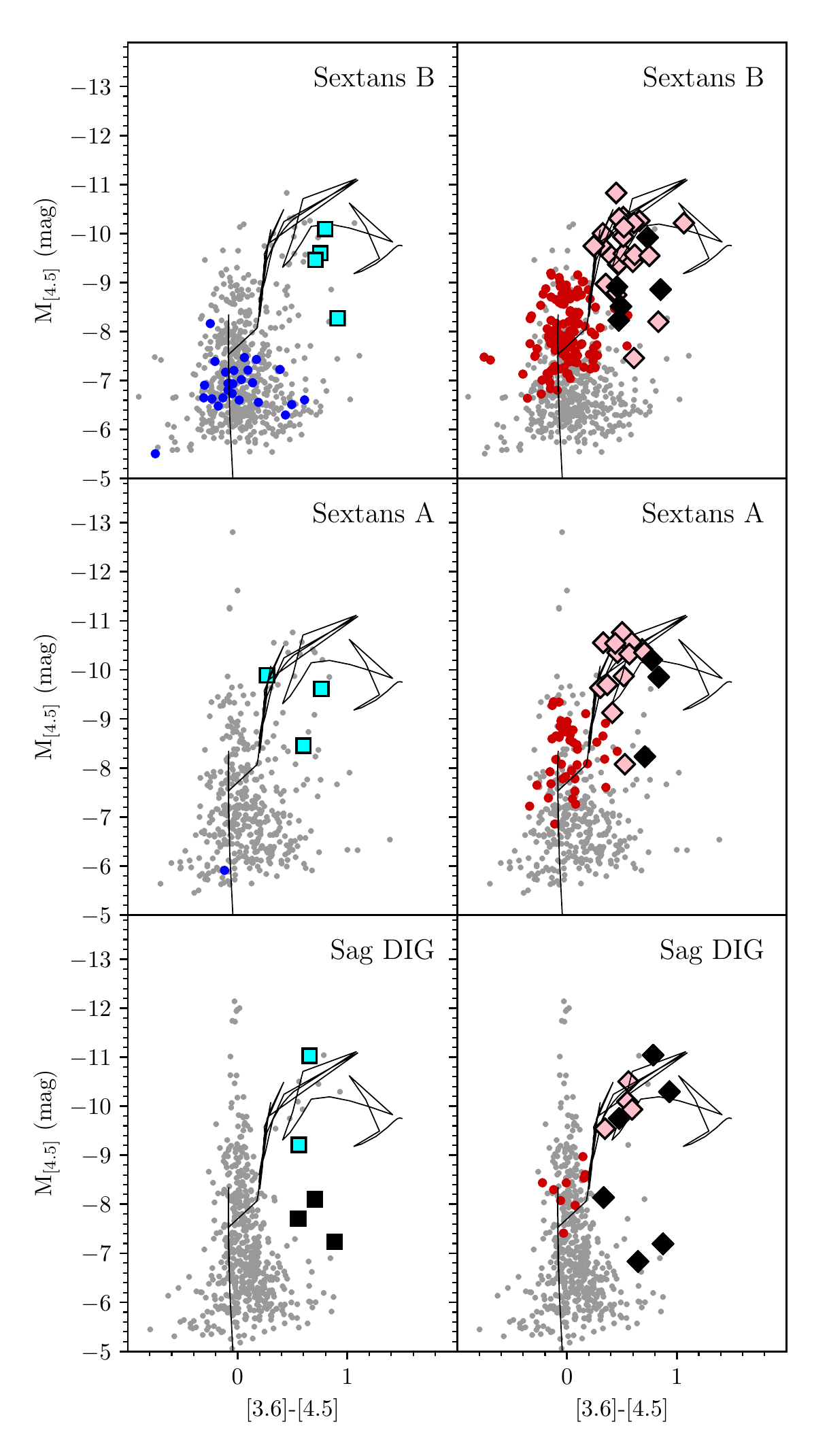}

    }
   \caption{{\it Spitzer} CMDs with M (blue) and C (red) stars
     overplotted. Symbols are the same as Figure~\ref{fig:hcmd}. The
     isochrone (thin black line) is the same one plotted in
     Figure~\ref{fig:not}, with $\log({\rm age}) = 8.8$. \label{fig:scmd}}
\end{figure*}

To minimize contamination, we use data from the SMC as a guide. The
Surveying the Agents of Galaxy Evolution (SAGE) program targeted
hundreds of sources in both the LMC and the SMC with the InfraRed
Spectrograph (IRS) onboard {\it Spitzer} \citep{Kemper+2010} and
\citet{Ruffle+2015} compared the spectroscopically-classified sources
to photometric classifications in the SMC. While their near-IR filters
are different from ours (they use $J$ and $K_{\rm S}$), it is clear
from Figure~13 in \citet{Ruffle+2015} that dusty AGB stars follow a
branch that extends from the TRGB to red colors and faint
magnitudes. A similar branch is evident in our near-IR CMDs
(Fig.~\ref{fig:hcmd}). YSOs, PNe, and post-AGB stars, on the other
hand, tend to be faint and blue in the near-IR. We therefore flag
sources to the left of the solid line in Figure~\ref{fig:hcmd} as
possible contaminants. The slope of the contaminant line was
determined using the expected direction of circumstellar extinction
from \citep{Groenewegen+2012}.

Another source of contamination is from red supergiants (RSGs), which
have similar infrared properties to AGB stars. RSGs are not easily
distinguished with this dataset, though they do tend to be warmer than
AGB stars, and thus fall towards the lower right end of the M star
sequence in Figure~\ref{fig:modccd}. A sample of 6 confirmed RSGs in
IC\,10 from \citet{Britavskiy+2015} fall below the knee in
Figure~\ref{fig:modccd}, with some even falling within the foreground
sequence. We expect the number of RSGs in our AGB sample to be small,
given their comparative rarity. This is discussed more in
\S\ref{sec:mdust}.

\section{Discussion}
\label{sec:dust}

\subsection{Dusty C Stars}
\label{sec:cdust}

We identify 120 dust-producing carbon stars, almost twice the number
detected via 2-epoch variability in
Paper\,II. Figure~\ref{fig:scmd} shows the \spitzer\ color-magnitude
diagrams for sources detected in the \hst\ images. These CMDs are
significantly cleaner than those presented by Paper~I because the
contamination from extended background sources is substantially
reduced by including the high-resolution \hst\ data. The relative
positions of dusty C and M stars in the \spitzer\ CMD are similar,
though the dusty C stars tend to be more tightly concentrated than M
stars.

Sag\,DIG and Sextans\,A are the two most metal-poor galaxies in our
sample, with gas-phase ISM metallicities more than an order of
magnitude below solar, suggesting that even the youngest stars are
very metal-poor.  Yet, both galaxies show a sizeable population of
dust-producing C stars, with very similar colors to those seen in more
metal-rich galaxies. For C stars, the $[3.6]-[4.5]$ color is
approximately proportional to the dust-production rates
\citep[Fig.~\ref{fig:grams};][]{Riebel+2012,Srinivasan+2016,Sloan+2016}. The
similar colors of the C stars across our sample therefore suggest that
dust masses are similarly at all metallicities. The high efficiency of
the third dredge up at low metallicity \citep[e.g.,][]{Karakas+2002}
appears to provide metal-poor C stars with plenty of material for dust
condensation.

Several dusty sources identified as C stars by the \hst\ color
definitions from \S\ref{sec:cmclass} are flagged as possible
contaminants based on their positions in Figure~\ref{fig:hcmd}. This
includes some of the reddest sources identified in the survey (see
Fig.~\ref{fig:scmd}). There is a strong possibility that these sources
are C-rich PNe or post-AGB stars. In the Magellanic Clouds, the
post-AGB stars tend to be brighter and redder than PNe at
4.5~\micron\ \citep{Ruffle+2015, Jones+2017}. For this reason, we
favor the possibility that the contaminants around $M_{[4.5]} \sim
-10$~mag are post-AGB stars, including the reddest object in IC\,10.

The reddest (non-potential-contaminant) carbon stars in our
  sample have $[3.6]-[4.5] \sim 1$~mag, which corresponds to a
  dust-production rate of $\log \dot{M}_{\rm dust} = -9$ to $-8$
  $[M_\odot\,{\rm yr}^{-1}$], according to the SMC relationship in
  Figure~\ref{fig:grams}.

\subsection{Dusty M stars}
\label{sec:mdust}

We find a total of 26 dust-producing M-type candidates, comprising
1.2\% of the M star population. We list them in Table~\ref{tab:M}
along with their photometry and classification confidence. The reddest
among these have $[3.6]-[4.5] \sim 1$~mag, which corresponds to a
dust-production rate of $\log \dot{M}_{\rm dust} = -7.5$ to $-6.5$
$[M_\odot {\rm\, yr}^{-1}$] using the SMC relationship in
Figure~\ref{fig:grams}.

AGB stars remain O-rich M-type stars both at the low and high end of
the AGB mass range due either to insufficient dredge up or to HBB,
though the exact mass limits depend on the metallicity.  Most of the
M-type stars we identified in Section~\ref{sec:cmclass} are low-mass
AGB stars and are expected to produce only modest amounts of dust
\citep[e.g.][]{McDonald+2009, McDonald+2011b, Boyer+2015c}. The
significant IR excesses of the 26 M-type dusty stars identified here
suggests that they are instead HBB AGB stars and are therefore more
massive than their C-rich counterparts. Moreover, the dusty M-type
stars are mostly confined to within each galaxy's half-light radius
\citep{McConnachie2012}. This suggests they are more massive than the
dusty C stars, which are located throughout the spatial coverage of
the \hst\ data including in fields that are entirely outside the
half-light radii (i.e., Fields A in NGC\,147 and Sag\,DIG). Age
gradients are expected in dwarf star-forming galaxies
\citep[e.g.,][see also Paper III]{AparicioTikhonov2000,Hidalgo+2013}
and this is expected from models of the effects of feedback and
stellar migration \citep[e.g.,][]{Stinson+2009,
  El-Badry+2016}. Overall, we identify about 1\% of the AGB stars as
massive M type, similar to the fraction seen in the SMC
\citep{Boyer+2011}.  If confirmed, their high masses and low
metallicities make them the closest known analogs to high-redshift
dusty AGB stars.

\begin{deluxetable*}{ccrcccccr}
  \tabletypesize{\scriptsize}
  \tablecolumns{9}
    \tablecaption{Dusty M stars \label{tab:M}}
    \tablehead{
      \colhead{Galaxy} &
      \colhead{DUSTiNGS} &
      \colhead{RA} &
      \colhead{F127M} &
      \colhead{F139M} &
      \colhead{F153M} &
      \colhead{[3.6]} &
      \colhead{[4.5]} &
      \colhead{Note}\\
      &
      \colhead{ID} &
      \colhead{Dec} &
      \colhead{(mag)} &
      \colhead{(mag)} &
      \colhead{(mag)} &
      \colhead{(mag)} &
      \colhead{(mag)} &}
    \startdata
    N 147 & 103322 & 00h33m09.79s & $18.849 \pm 0.005$ & $18.718 \pm 0.006$ & $18.293 \pm 0.005$ & $16.59 \pm 0.05$ & $16.19 \pm 0.06$ & H\\
     && $+$48d29m09.28s & & & & &  & \\
    N 147 & 112780 & 00h33m05.61s &   $21.006 \pm 0.021$ & $20.524 \pm 0.018$ & $19.992 \pm 0.015$ & $15.28 \pm 0.03$ & $14.70 \pm 0.03$ & Lx \\
    && $+$48d28m47.43s & & & & &  & \\
    IC 10 & 102032 & 00h20m14.98s &   $18.883 \pm 0.006$ & $18.768 \pm 0.006$ & $18.181 \pm 0.005$ & $17.10 \pm 0.05$ & $16.80 \pm 0.05$ & H\\
    && $+$59d21m01.16s & & & & &  & \\
    IC 10 & 105880 & 00h20m12.38s & $22.437 \pm 0.068$ & $21.889 \pm 0.052$ & $21.221 \pm 0.037$ & $15.36 \pm 0.05$ & $14.88 \pm 0.06$ & L\\
    && $+$59d19m41.95s  & & & & &  & \\
    IC 10 & 105975 & 00h20m12.30s &   $15.666 \pm 0.001$ & $15.518 \pm 0.001$ & $14.955 \pm 0.001$ & $13.16 \pm 0.03$ & $12.84 \pm 0.03$ & H\\
    && $+$59d20m42.09s & & & & &  & \\
    IC 10 & 107349 & 00h20m11.39s &   $19.530 \pm 0.015$ & $19.278 \pm 0.017$ & $18.821 \pm 0.017$ & $16.68 \pm 0.10$ & $16.06 \pm 0.07$ & L\\
    && $+$59d19m03.87s & & & & &  & \\
    IC 10 & 112431 & 00h20m07.99s &   $17.275 \pm 0.002$ & $17.727 \pm 0.003$ & $16.856 \pm 0.002$ & $14.91 \pm 0.05$\tablenotemark{a} & $14.73 \pm 0.04$ & Hx\\
    && $+$59d19m31.74s & & & & &  & \\
    IC 10 & 118138 & 00h20m04.07s &   $17.420 \pm 0.003$ & $17.764 \pm 0.004$ & $17.029 \pm 0.003$ & $14.30 \pm 0.03$ & $14.08 \pm 0.03$ & Hx\\
    && $+$59d19m30.50s & & & & &  & \\
    IC 10 & 121876 & 00h20m01.50s &   $18.724 \pm 0.005$ & $18.516 \pm 0.005$ & $18.010 \pm 0.004$ & $15.18 \pm 0.03$ & $14.32 \pm 0.03$ & Hx\\
    && $+$59d20m02.05s & & & & &  & \\
    IC 10 & 122923 & 00h20m00.81s &   $18.816 \pm 0.005$ & $19.590 \pm 0.010$ & $18.596 \pm 0.006$ & $16.40 \pm 0.03$ & $16.04 \pm 0.05$ & H\\
    && $+$59d20m32.04s & & & & &  & \\
    IC 10 & 96092 & 00h20m19.00s &   $20.783 \pm 0.019$ & $20.413 \pm 0.020$ & $19.955 \pm 0.018$ & $14.35 \pm 0.02$ & $13.51 \pm 0.03$ & Lx\\
    && $+$59d16m39.47s & & & & &  & \\
    IC 10 & 98013 & 00h20m17.75s &  $20.568 \pm 0.016$ & $19.930 \pm 0.012$ & $19.235 \pm 0.009$ & $14.18 \pm 0.02$ & $13.62 \pm 0.03$ & Lx\\
    && $+$59d16m18.22s & & & & &  & \\
    IC 10 & 107616 & 00h20m11.30s &   $19.040 \pm 0.006$ & $18.798 \pm 0.006$ & $18.259 \pm 0.005$ & $16.36 \pm 0.05$ & $15.82 \pm 0.06$ & H\\
    && $+$59d17m17.95s & & & & &  & \\
    IC 10 & 108360 & 00h20m10.81s &   $20.113 \pm 0.012$ & $19.591 \pm 0.010$ & $19.030 \pm 0.008$ & $15.48 \pm 0.03$ & $14.85 \pm 0.03$ & L\\
    && $+$59d16m12.22s & & & & &  & \\
    IC 10 & 117402 & 00h20m04.72s &   $20.173 \pm 0.012$ & $19.937 \pm 0.012$ & $19.315 \pm 0.009$ & $17.17 \pm 0.07$ & $16.75 \pm 0.07$ & H\\
    && $+$59d17m16.20s & & & & &  & \\
    IC 10 & 120247 & 00h20m02.74s &   $22.315 \pm 0.063$ & $21.526 \pm 0.042$ & $20.642 \pm 0.024$ & $15.74 \pm 0.03$ & $14.78 \pm 0.03$ & H\\
    && $+$59d17m18.08s & & & & &  & \\
    IC 10 & 120468 & 00h20m02.58s &   $21.510 \pm 0.033$ & $20.860 \pm 0.025$ & $20.074 \pm 0.017$ & $15.20 \pm 0.02$ & $14.69 \pm 0.03$ & Hx\\
    && $+$59d16m52.46s & & & & &  & \\
    Sex B & 84519 & 10h00m03.61s &   $22.688 \pm 0.078$ & $22.249 \pm 0.068$ & $21.659 \pm 0.053$ & $18.42 \pm 0.07$ & $17.50 \pm 0.05$ & L\\
    && $+$05d18m55.29s & & & & &  & \\
    Sex B & 109067 & 09h59m55.70s &   $20.917 \pm 0.020$ & $20.443 \pm 0.018$ & $19.937 \pm 0.016$ & $16.48 \pm 0.03$ & $15.68 \pm 0.06$ & Lx\\
    && $+$05d19m53.04s & & & & &  & \\
    Sex B & 116156 & 09h59m53.43s &   $21.706 \pm 0.059$ & $21.232 \pm 0.043$ & $20.644 \pm 0.034$ & $16.93 \pm 0.05$ & $16.18 \pm 0.05$ & L\\
    && $+$05d18m51.76s & & & & &  & \\
    Sex B & 123385 & 09h59m51.09s &   $21.060 \pm 0.076$ & $20.822 \pm 0.073$ & $20.463 \pm 0.061$ & $17.02 \pm 0.06$ & $16.31 \pm 0.05$ & L\\
    && $+$05d19m26.61s & & & & &  & \\
    Sex A & 90034 & 10h10m59.54s &   $17.860 \pm 0.004$ & $17.835 \pm 0.004$ & $17.404 \pm 0.003$ & $16.21 \pm 0.03$ & $15.94 \pm 0.05$ & H\\
    && $-$04d40m58.69s  & & & & &  & \\
    Sex A & 91324 & 10h10m59.08s &   $22.921 \pm 0.098$ & $22.707 \pm 0.103$ & $21.936 \pm 0.068$ & $17.98 \pm 0.05$ & $17.37 \pm 0.05$ & L\\
    && $-$04d43m59.30s & & & & &  & \\
    Sex A & 94328 & 10h10m58.03s &   $22.413 \pm 0.065$ & $21.965 \pm 0.057$ & $21.112 \pm 0.038$ & $16.98 \pm 0.05$ & $16.22 \pm 0.05$ & Lx\\
    && $-$04d43m04.22s & & & & &  & \\
    Sag & 44334 & 19h29m57.94s &   $18.034 \pm 0.004$ & $17.723 \pm 0.004$ & $17.256 \pm 0.003$ & $14.83 \pm 0.03$ & $14.17 \pm 0.03$ & H\\
    && $-$17d40m17.43s & & & & &  & \\
    Sag & 45478 & 19h29m57.29s &   $21.245 \pm 0.050$ & $20.966 \pm 0.053$ & $20.559 \pm 0.041$ & $16.55 \pm 0.06$ & $15.99 \pm 0.08$ & L\\
    && $-$17d40m11.28s & & & & &  & 
    \enddata
    
    \tablecomments{\ Near-IR {\it HST} and {\it Spitzer} magnitudes
      for candidate M-type, dusty AGB stars. Magnitudes are corrected
      for extinction (see text). This includes stars in the M-type
      region of Figure~\ref{fig:modccd} and with $>$4$\sigma$ excess
      in $[3.6]-[4.5]$ color. The {\it Spitzer} magnitudes are from
      the \dustings\ survey \citep{Boyer+2015a,Boyer+2015b}. The {\it
        Note} column marks M type AGB stars identified with
      high-confidence (H; HST colors indicate water absorption) and
      low-confidence (L; star is near the border of C and M stars, is
      near the contamination cutoff in Fig.~\ref{fig:hcmd}, or is
      affected by crowding). An ``x'' denotes a \dustings\ x-AGB
      variable (those with $>$3$\sigma$ variability between two {\it
        Spitzer} epochs, i.e., a likely AGB star). Non-x-AGB stars may
      also be variable, but were observed by \dustings\ with an unfavorable
      cadence. }
    
    \tablenotetext{a}{\ The IRAC magnitudes listed for 112431 are from \dustings\ epoch~2 because the star is not as red in epoch~1 ($<$4-$\sigma$ excess).\citep{Boyer+2015a}.}

\end{deluxetable*}


The brightest among the massive AGB candidates may instead be RSG
stars, which are expected to destroy all or most of their dust when
they explode as supernovae \citep{Lakicevic+2015, Temim+2015}.  It can
be difficult to distinguish M-type AGB stars from RSGs -- luminosity
is often used as a diagnostic, with the classical AGB limit near
$M_{\rm bol} = -7.1$~mag. However, HBB AGB stars have been known to
surpass this luminosity \citep[e.g.,][]{Garcia-Hernandez+2009}. Water
absorption is another possible diagnostic; \citet{Lancon+2007} notes
that water absorption in the near-IR is present in Galactic RSGs with
cool effective temperatures ($T_{\rm eff} < 3100$~K for $\log g =
-1$), but is weak compared to the strength of water features in Mira
variables. \citet{Messineo+2014} use the \citet{Rayner+2009} IRTF
spectral library to compute a water absorption index and find that the
index is large for all Mira-variable AGB stars, and small for
semiregular variable AGB stars and RSGs. It follows that the H$_2$O
absorption sequence in the \hst\ CCD (Figs.~\ref{fig:modccd} and
\ref{fig:ccds}) can be used as a diagnostic to separate probable AGB
and RSG stars. M-type stars on the water absorption sequence are
probable AGB stars, while those at or below the `knee' in
Figure~\ref{fig:modccd} may be either less-evolved AGB stars or
RSGs. Variability is an additional diagnostic; stars that were
identified as variable in \dustings\ are probable AGB stars, given
their large infrared amplitudes.

Based on their H$_2$O and variability signatures, IC\,10 has several
candidate dusty M-type AGB stars likely undergoing HBB: ten identified
with high confidence and five with low confidence
(Table~\ref{tab:M}). There are two additional high confidence dusty
M-type candidates in NGC\,147 (\#103322) and in Sextans\,A
(\#90034). Low-confidence dusty M-type stars either show weak H$_2$O
absorption, fall near the contamination cutoff in
Figure~\ref{fig:hcmd}, or fall on the dividing line between C- and
M-type stars in Figure~\ref{fig:ccds}, and thus a C-rich chemistry
cannot be ruled out.  Other low-confidence dusty M-type stars are
affected by crowding, which may influence the near-IR colors by up to
0.5~mag (as measured by Dolphot; Section~\ref{sec:data}); these
include \#91324 and \#94328 in Sextans\,A, \#45478 in Sag\,DIG, and
\#96092 in IC\,10.

The candidates in Sextans\,A and Sag\,DIG (especially \#44334 and
\#90034) are of particular interest. The gas-phase ISM metallicities
(and thus the metallicities of even the most massive stars) are low
for both galaxies ($12+\log({\rm O/H}) \lesssim 7.5$) suggesting that
stars with primordial abundances can still form a significant dust
mass. Whether this occurs via single-star evolution or through a more
exotic avenue remains unclear. On the other hand, the high near-IR
luminosities of both stars suggests they may in fact be RSGs, though
\#90034 in Sextans\,A does show evidence for water absorption and Star
\#44334 is a known long-period variable
\citep[$P=950$~days;][]{Whitelock+2017}. Both of these characteristics
point to an AGB nature. Three additional confirmed O-rich stars in
Sag\,DIG that were identified by \citet{Momany+2014} are included in
our near-IR HST coverage. These 3 stars have very red F606W$-$F814W
colors, suggesting substantial circumstellar dust, but we do not
detect dust around these stars. Their nature is unclear.

We have too small a sample to draw definitive conclusions, but there
is nonetheless no hint of increased dust production at higher
metallicities.  The dusty M-type stars have remarkably similar properties in
the {\it Spitzer} data, including colors, magnitudes, and pulsation
amplitudes.  Spectroscopy can definitively classify these
sources, especially at wavelengths covering ice features common in
YSOs (3--15~\micron) or in the optical where the \ion{Li}{1}
6707-\AA\ and 8126-\AA\ and \ion{Rb}{1} 7800-\AA\ lines indicate HBB
in AGB stars.

\subsubsection{Super AGB Stars}
\label{sec:SAGB}

Four dusty M-type stars may be examples of super AGB stars: stars just
under the mass transition for becoming a high-mass star in which core
nucleosynthesis proceeds to iron before a core collapse
(8--12~$M_\odot$). Super AGB stars ignite carbon in their cores,
leaving behind an ONe white dwarf \citep{Siess2007}, though some super
AGB stars may ultimately explode as electron-capture SNe
\citep[e.g.,][]{Doherty+2015,Doherty+2017}.

Models from \citet{Doherty+2015} suggest that metal-poor super AGB
stars can become as much as a magnitude brighter than the classical
AGB limit, which is near $-9.4$~mag in F153M. Star \#105975 in IC\,10
has $M_{\rm F153M} = -9.9$~mag and is our best super AGB candidate. Of
course, there is the possibility this star is a red supergiant, but
its strong water absorption signature (F127M$-$F139M = 0.07~mag)
suggests otherwise.

In addition, star \#90034 in Sextans\,A and stars \#112431 and \#118138 in
IC\,10 are within $\approx$1~mag of the classical AGB limit and also
show significant water absorption and dust excess. The two IC\,10
stars also showed variability in \dustings. It is thus very likely that
these are massive AGB stars, but whether they are super AGB stars
remains to be seen.

Additional super-AGB candidates are the 9 infrared variables without
\hst\ counterparts (Table~\ref{tab:nohst}). These are among the
reddest objects in our sample, and include examples in 3 galaxies:
IC\,10, Sextans\,A, and Sextans\,B. The {\it Spitzer} colors and
magnitudes are similar to the progenitor of SN2008S
\citep{Prieto+2008} and to other intermediate-luminosity IR transients
\citep[e.g.,][]{Kasliwal+2017}.

Detection of Li or Rb in these stars would confirm that they are
indeed HBB AGB stars, adding to a very small collection of massive
and/or super AGB candidates in the Magellanic Clouds \citep{Plez+1993,
  Smith+1995, Garcia-Hernandez+2009, Groenewegen+2009,
  GroenewegenSloan2017} and in IC\,1613 \citep{Menzies+2015}.
However, we note that IC\,10 is in the direction of the Galactic
Plane, so finding a foreground AGB star is conceivable. IC\,10 star
\#105975 is in the Gaia Data Release 1 catalog, but has no reported
parallax because it is too faint for the Tycho-2 catalog \citep[$G$ =
  20.06~mag;][]{Gaia}.

\subsection{Implications for the Dust Budget}

Based on the results in the SMC \citep[see
  Fig.~\ref{fig:grams};][]{Srinivasan+2016}, O-rich AGB stars redder
than $[3.6]-[4.5] \approx 0.1$~mag have dust-production rates that are
an order of magnitude or more larger than C-rich stars {\it with the
  same color.} In our fields, the dusty C-rich stars outnumber the
dusty O-rich sources by factors of about 2--10, but the total dust
mass may be more evenly split between silicates and carbon grains if
the M-type dust-production rates behave similarly to those in the
SMC. The rarity of dust-producing M-type stars makes the balance
between O-rich and C-rich dust highly stochastic. Observations with
the James Webb Space Telescope will provide more reliable estimates of
the dust-production rates for entire stellar populations in dwarf
galaxies well beyond the Local Group \citep{Boyer2016}.

\section{Conclusions}
\label{sec:conclusions}

The \dustings\ survey identified hundreds of candidate dust-producing
AGB stars in nearby galaxies spanning a broad metallicity range,
finding that AGB stars can form dust at metallicities as low as 0.6\%
solar (Paper~II). However, that survey was unable to identify the
stars' spectral types, and thus could not infer whether the primary
dust species produced is carbon- or oxygen-rich. We surveyed six
\dustings\ galaxies using medium-band filters on \hst's WFC3/IR to
identify AGB spectral types. These filters sample the near-IR CN and
C$_2$ features in carbon stars and the H$_2$O feature in oxygen-rich
stars, providing an effective tool for separating C-, M-, and K-type
stars. The galaxies surveyed here include 99 candidate dusty stars and
span 1~dex in metallicity ($-2.1 < {\rm [Fe/H]} < -1.1$). Altogether,
we identify 908 C stars and 2120 M stars among these galaxies; 13.2\%
and 1.2\% of these, respectively, show evidence for dust production
(see Table~\ref{tab:data}). Our conclusions are as follows:

\begin{itemize}[leftmargin=*]
\item[--] Most of the dusty AGB candidates identified in Paper II are
  confirmed here to be C-type AGB stars, confirming that C-rich AGB stars can
  form dust in very metal-poor environments.
\item[--] We find 26 dusty M-type stars among our sample, including in
  our most metal-poor galaxies (Sextans\,A and Sag\,DIG).  These stars
  are very likely metal-poor given the low ISM gas-phase metallicities
  of these galaxies ($12+\log{\rm (O/H)} = 7.2$--$7.5$). Finding
  dust-producing M-type stars at these metallicities is a surprise;
  unlike C stars, M-type stars do not to produce their own condensable
  material so dust production is not expected at low metallicity.
  Given their dust excess and their central locations within each
  galaxy, it is likely that these are massive AGB stars. This makes
  them the closest known analogs to AGB stars at high redshift that
  may be contributing to galaxy dust budgets as early as 30~Myr after
  they form.
\item[--] The brighest dusty M-type star in our sample (in IC\,10) exceeds
  the classical AGB limit, but its strong water absorption signature
  (as evident in the F127M$-$F139M and F139M$-$F153M colors) suggests
  that it is an AGB star rather than a more massive red supergiant. This star may
  be an example of an (exceptionally rare) super AGB star ($M_{\rm i}
  =8$--$12~M_\odot$), which will end its life either as an ONe white
  dwarf or an electron capture supernova.
\end{itemize}

\acknowledgements Support for HST program GO-14073 was provided by
NASA through a grant from the Space Telescope Science Institute, which
is operated by the Association of Universities for Research in
Astronomy, Incorporated, under NASA contract NAS5-26555. Analysis for
this work made use of Astropy \citep{astropy}. We thank the referee
for suggestions that improved the manuscript.

\bibliographystyle{aasjournal.bst}
\bibliography{myrefs}

\end{document}